\documentclass[a4paper,11pt]{article}

\usepackage{jcappub}
\usepackage{graphicx}
\usepackage{amsmath,amsfonts,amssymb}
\usepackage{hyperref}
\usepackage{color}

\def\lsim{\mathrel{\raise.3ex\hbox{$<$\kern-.75em\lower1ex\hbox{$\sim$}}}}
\def\gsim{\mathrel{\raise.3ex\hbox{$>$\kern-.75em\lower1ex\hbox{$\sim$}}}}

\newcommand{\be}{\begin{equation}}
\newcommand{\ee}{\end{equation}}
\newcommand{\bea}{\begin{equation}\begin{aligned}}
\newcommand{\eea}{\end{aligned}\end{equation}}
\newcommand{\td}{{\rm d}}

\begin{document}
\begin{flushright}
KCL-PH-TH/2020-40, CERN-TH-2020-129
\end{flushright}

\title{Updated predictions for gravitational waves produced in a strongly supercooled phase transition}

\author[1,2,3]{John~Ellis,}
\author[1,4]{Marek~Lewicki,}
\author[1,3]{and Ville~Vaskonen}
\affiliation[1]{Department of Physics, King's College London, Strand, London WC2R 2LS, UK}
\affiliation[2]{Theoretical Physics Department, CERN, Geneva, Switzerland}
\affiliation[3]{National Institute of Chemical Physics \& Biophysics, R\"avala 10, 10143 Tallinn, Estonia}
\affiliation[4]{Faculty of Physics, University of Warsaw, ul.\ Pasteura 5, 02-093 Warsaw, Poland}
\emailAdd{john.ellis@cern.ch}
\emailAdd{marek.lewicki@kcl.ac.uk}
\emailAdd{ville.vaskonen@kcl.ac.uk}

\abstract{We update predictions for the gravitational wave (GW) signal from a strongly supercooled phase transition in an illustrative classically conformal U(1)$_{B-L}$ model. We implement $\propto \gamma^2$ scaling of the friction on the bubble wall and update the estimates for the efficiency factors for GW production from bubble collisions and plasma-related sources. We take into account the fact that a small decay rate of the symmetry-breaking field may lead to brief matter-dominated  era after the transition, as the field oscillates around its minimum before decaying. We find that a strong bubble collision signal occurs in a significant part of the parameter space, and that the modified redshift of the modes that re-enter the horizon during the matter-dominated period generates a characteristic tilted `plateau' in the spectrum. The GW spectrum in this model would be detectable in the low-frequency range, e.g., by LISA, and in the mid-frequency range, e.g., by AION/MAGIS and AEDGE, and in the high-frequency range by LIGO and ET. The peak frequency of the signal is limited from below by collider constraints on the mass of the U(1)$_{B-L}$ gauge boson, while at high frequencies the slow decay of the scalar field and the resulting matter-dominated era diminishes the GW signal.}

\maketitle

\section{Introduction}

The electroweak phase transition is not predicted to be first order within the Standard Model (SM). However, a first-order phase transition is a common feature in many extensions of the SM, including models with higher-dimensional multi-Higgs interactions, additional scalar fields and extended gauge groups. Historically, the possibility of a first-order phase transition attracted interest because it facilitated electroweak baryogenesis~\cite{Kuzmin:1985mm,Cohen:1993nk,Riotto:1999yt,Morrissey:2012db}. However, the direct detection of gravitational waves (GWs) by the LIGO and Virgo Collaborations~\cite{Abbott:2016blz} has recently stimulated widespread interest in the potential observability of GWs~\cite{Witten:1984rs} produced in such a transition~\cite{Kamionkowski:1993fg,Grojean:2006bp,Espinosa:2008kw,Dorsch:2014qpa,Jaeckel:2016jlh,Jinno:2016knw,Chala:2016ykx,Chala:2018opy,Artymowski:2016tme,Hashino:2016xoj,Vaskonen:2016yiu,Dorsch:2016nrg,Beniwal:2017eik,Baldes:2017rcu,Marzola:2017jzl,Kang:2017mkl,Iso:2017uuu,Chala:2018ari,Bruggisser:2018mrt,Megias:2018sxv,Croon:2018erz,Alves:2018jsw,Baratella:2018pxi,Angelescu:2018dkk,Croon:2018kqn,Brdar:2018num,Beniwal:2018hyi,Breitbach:2018ddu,Marzo:2018nov,Baldes:2018emh,Prokopec:2018tnq,Fairbairn:2019xog,Helmboldt:2019pan,Dev:2019njv,Jinno:2019jhi,Ellis:2019flb,Jinno:2019bxw,Azatov:2019png,Hindmarsh:2019phv,vonHarling:2019gme,DelleRose:2019pgi}. Another interesting possibility is that a first-order phase transition may have played a role in the generation of an intergalactic magnetic field~\cite{Vachaspati:1991nm,Sigl:1996dm,Tevzadze:2012kk,Ellis:2019tjf}.

A primary ingredient in the calculation of the GW spectrum arising from a first-order phase transition is the calculation of the velocities of walls of bubbles of the true vacuum. Over the years many groups have contributed to the progress in this notoriously difficult calculation~\cite{Moore:1995si,Moore:1995ua,Bodeker:2009qy,Kozaczuk:2015owa,Bodeker:2017cim,Dorsch:2018pat,Mancha:2020fzw,Hoeche:2020rsg}. In particular, there has recently been a significant advance in the estimation of the scaling of the friction, $P$, encountered by the wall with its Lorentz factor, with the result that $P \propto \gamma^2$~\cite{Hoeche:2020rsg}, to be contrasted with the previous estimate~\cite{Bodeker:2017cim} that $P \propto \gamma$.

This change in the scaling law has a crucial impact on the amount of energy that is transferred into the plasma instead of the scalar field gradients, which tends to reduce the importance of bubble wall collisions as a source of GWs~\cite{Ellis:2019oqb}. However, we show here that this modification does not necessarily change qualitatively the GW predictions found previously in models featuring a very strong first-order phase transition. To illustrate this point, we revisit the classically conformal $U(1)_{B-L}$ extension of the SM~\cite{Jinno:2016knw,Iso:2017uuu,Marzo:2018nov,Ellis:2019oqb}, and show that there is no dramatic change in the regions of the parameter space where a strong GW signal from bubble collisions is predicted.

Also, we highlight a feature in the GW spectrum that can appear in many models predicting such a strong transition. As the strength of the transition increases, the decay width of the field involved decreases and, as a result, the reheating after the transition is not necessarily instantaneous. It can instead be preceded by a period during which the field oscillates around its minimum, effectively as during a matter-dominated epoch~\cite{Allahverdi:2020bys}. This period is reflected in the GW spectrum, because it modifies the redshifting history~\cite{Barenboim:2016mjm}. This effect produces a characteristic tilted plateau-like feature in the part of the spectrum due to super-horizon modes, which could in principle be used by GW detectors to probe the field decay rate directly.

The structure of this paper is as follows. In Section~\ref{sec:model} we review the classically conformal ${\rm U}(1)_{B-L}$ model that we study, and in Section~\ref{sec:PT} we discuss the phase transition in this model, delineating in Fig.~\ref{fig:T} the region where this transition is strongly first-order. Then, in Section~\ref{sec:gamma} we consider the Lorentz factor of the bubble wall, commenting that the bubbles collide before they reach terminal velocity in much of the region where the transition is strong, and updating the estimate for the efficiency factor for GW production from bubble collisions. In Section~\ref{sec:gws}  we present the calculation of the GW signal and in Section~\ref{sec:scaling} we consider the impact on the signal of a matter-dominated period, displaying the plateau feature that is a signal of the slow decay of the ${\rm U}(1)_{B-L}$-breaking scalar field. Finally, in Section~\ref{sec:results} we discuss the observability of the GW signal from the phase transition by LIGO~\cite{TheLIGOScientific:2014jea,Thrane:2013oya,TheLIGOScientific:2016wyq} and the future GW detectors LISA~\cite{Bartolo:2016ami,Caprini:2019pxz}, AION~\cite{Badurina:2019hst}, MAGIS~\cite{Graham:2016plp,Graham:2017pmn}, AEDGE~\cite{Bertoldi:2019tck} and the Einstein Telescope~(ET)~\cite{Punturo:2010zz,Hild:2010id}, as summarised in Fig.~\ref{fig:SNR}. Finally Section~\ref{sec:concl} summarises our results and discusses their implications.

\section{Classically conformal ${\rm U(1)}_{B-L}$ model}
\label{sec:model}

Following Ref.~\cite{Marzo:2018nov}, we describe briefly the scalar potential and the symmetry-breaking pattern in a classically conformal ${\rm U}(1)_{B-L}$ extension of the SM~\cite{Iso:2009ss}. In addition to the SM particle content, this model includes the ${\rm U}(1)_{B-L}$ gauge boson $Z'$, three right-handed neutrinos that cancel the ${\rm U}(1)_{B-L}$ anomaly, and a complex scalar $\phi = (\varphi+iG)/\sqrt{2}$ with $B-L$ charge $+2$, responsible for the breaking of the ${\rm U}(1)_{B-L}$ symmetry in vacuum. At tree level, the scalar potential of this model is
\be
V = \lambda_H (H^\dagger H)^2 + \lambda_\phi (\phi^\dagger \phi)^2 - \lambda_p (H^\dagger H)(\phi^\dagger \phi) \,,
\ee
where $H$ is the Higgs doublet and $\lambda_H,\lambda_\phi,\lambda_p>0$. 

For values of the $B - L$ gauge coupling $g_{B-L} \sim 0.1$, collider searches give a lower bound for the $B - L$ gauge boson mass: $m_{Z'}\gsim 4$\,TeV~\cite{Aaboud:2017buh,Escudero:2018fwn}. This can be converted into a lower bound on the $\varphi$ vacuum expectation value $\langle \varphi \rangle = w$ though $m_{Z'} = 2g_{B-L} w$, giving $w > \mathcal{O}(1\,{\rm TeV}) \gg v \approx 246\,$GeV. In this case the symmetry breaking occurs first along the $\varphi$ direction. The one-loop effective potential~\footnote{We take this opportunity to point out that the potential, and hence our results, are subject to some modification in a fully gauge-invariant formalism~\cite{Wainwright:2011qy,Chiang:2017zbz}.} along this direction is given by
\be \label{Veff1}
V_{\rm eff} = \frac{\lambda_\phi(t)}{4} \varphi^4 + \frac{T^4}{2\pi^2} \sum_j k_j J_T \left(m_j(\varphi)^2 + \Pi_j(T) \right)  \,.
\ee
The first term in Eq.~\eqref{Veff1} is the $T=0$ one-loop renormalisation-group (RG)-improved scalar potential with the RG scale chosen to $t=\log(\varphi/\mu_0)$, where $\mu_0$ is a reference scale for which we take the top quark mass $\mu_0 = 173\,$GeV. The second term accounts for one-loop finite-temperature effects. The sum over $j$ includes the $B - L$ gauge boson, the right-handed neutrinos, the scalar boson $\varphi$ and the Goldstone boson $G$, and the factors $k_j$ are their numbers of intrinsic degrees of freedom. The field-dependent masses for the U$(1)_{B-L}$ gauge boson $Z'$, the right-handed neutrinos ($i=1,2,3$), the scalar boson $\varphi$ and the Goldstone boson $G$ are given by
\bea
& m_{Z'}(\varphi)^2 = 4 g_{B-L}(t)^2 \varphi^2\, \quad M_i(\varphi)^2 = Y_i(t)^2\varphi^2/2 \,, \\
& m_\varphi(\varphi)^2 = 3\lambda_\phi(t) \varphi^2\,, \quad m_G(\varphi)^2 = \lambda_\phi(t) \varphi^2 \,,
\eea
and their Debye masses are
\bea
& \Pi_{Z'}(T) = 4 g_{B-L}(t)^2 T^2\,, \\
& \Pi_\varphi(T) = \frac{T^2}{24}\left[ 24 g_{B-L}(t)^2 + 8 \lambda_\phi(t) + \sum_i Y_i(t)^2 \right] \,, \\
& \Pi_G(T) = \Pi_\varphi(T) \,.
\eea
The Yukawa couplings of the right-handed neutrinos to $\phi$ are denoted by $Y_i$. The thermal integral $J_T$ is defined as 
\be
J_T(x) = \int_0^\infty {\rm d}y\, y^2 \ln\left[1\mp e^{-\sqrt{x+y^2}}\right] \,,
\ee
with the negative sign for bosons and the positive sign for fermions.

The vacuum expectation value of $\varphi$ induces a negative mass term for the Higgs field, and the electroweak symmetry breaks in the same way as in the SM. However, if the $B-L$ breaking is delayed to temperatures below the QCD scale, the QCD phase transition happens first, inducing a linear term for the Higgs field, $\sum_j y_j \langle \psi_j \bar\psi_j \rangle h/\sqrt{2}$. The Higgs field then acquires a small non-zero expectation value. Therefore, for $T<T_{\rm QCD}$ the effective potential along the $\varphi$ direction becomes 
\be
V_{\rm eff}(T<T_{\rm QCD}) = - \frac{\lambda_p(t) v_{\rm QCD}^2}{4} \varphi^2 + V_{\rm eff}(T>T_{\rm QCD}) \,,
\ee
where $v_{\rm QCD} \simeq 0.1\,$GeV. After the transition the QCD and electroweak symmetries are restored as the decay of the vacuum energy reheats the plasma, and, again, the evolution proceeds as in the SM. In the following we focus on the U$(1)_{B-L}$-breaking phase transition. 

The model also includes kinetic mixing between the ${\rm U}(1)_{B-L}$ and hypercharge ${\rm U}(1)_{Y}$ gauge symmetries. The kinetic mixing term can be removed by a rotation and a rescaling of the gauge fields. After this the ${\rm U}(1)_{B-L}\times{\rm U}(1)_{Y}$ part of the covariant derivative is given by
\be
\mathcal{D}_\mu \supset  i g_Y q_Y B_Y^\mu + i (\tilde g q_Y + g_{B-L} q_{B-L}) B_{B-L}^\mu \,,
\ee
where the coupling $\tilde g$ parametrizes the kinetic mixing, $q_j$ denotes the charges, $g_j$ the gauge couplings and $B_j^\mu$ the gauge fields. For $w\gg v$ and $g_{B-L}\sim 0.1$ the effect of the kinetic mixing on the $Z'$ mass can be neglected and also the mixing angle between $Z$ and $Z'$ gauge bosons is tiny (see e.g. Ref.~\cite{Coriano:2015sea}). The kinetic mixing affects the running of the Higgs quartic coupling and can stabilize the electroweak vacuum. In the following we fix $\tilde{g} = -0.5$ at $t = \log(w/\mu_0)$, which was shown in Ref.~\cite{Marzo:2018nov} to keep $\lambda_h$ positive while retaining the perturbativity of the couplings up to the Planck scale for $g_{B-L}\lsim 0.42$. The phase transition dynamics, however, is not very sensitive to the value of $\tilde{g}$, which enters the one-loop effective potential only through the running of $g_{B-L}$.

We assume that the Majorana Yukawa couplings of the right-handed neutrinos can be neglected, $Y_i \ll g_{B-L}$. We fix the couplings $\lambda_\phi$ and $\lambda_p$ and at $t = \log(w/\mu_0)$ so that $\td V/\td \varphi=0$ at $\varphi=w$ and $\lambda_p = v^2/w^2$. The remaining free parameters are then the gauge coupling $g_{B-L}$ and the vacuum expectation value $w$, which we trade for the $Z'$ mass, $m_{Z'} = 2g_{B-L} w$. In the following we show the results as a function of these two parameters, with the chosen value of $g_{B-L}$ corresponding to the scale $t = \log(w/\mu_0)$.

\section{Phase transition}
\label{sec:PT}

We compute the details of the phase transition in a standard way, starting from the decay rate of the false vacuum due to thermal fluctuations~\cite{Linde:1981zj}:
\be
\Gamma(T) \simeq T^4\left(\frac{S_3}{2\pi T}\right)^{\frac{3}{2}} e^{-S_3/T}\,,	
\ee
where the action of the O$(3)$ symmetric field configuration is
\be
S_3 = 4\pi \int r^2 {\rm d}r \left[ \frac{1}{2}\left(\frac{{\rm d}\varphi}{{\rm d}r}\right)^2 + V_{\rm eff}(\varphi,T) \right] \, .
\ee
The decay is dominated by the classical path that is the solution of the equation of motion obtained from the above action,
\be
\frac{{\rm d}^2 \varphi}{{\rm d} r^2} + \frac{2}{r} \frac{{\rm d} \varphi}{{\rm d} r} = \frac{{\rm d} V_{\rm eff}}{{\rm d} \varphi}\, ,
\ee
with boundary conditions ${\rm d}\varphi/{\rm d}r=0$ at $r=0$ for the solution to be regular at the bubble centre, and $\varphi\to0$ at $r\to\infty$ to describe the initial false vacuum background far from the bubble.

As shown in Ref.~\cite{Ellis:2018mja}, taking particular care of the subsequent evolution of bubbles is crucial for very strong transitions. The probability that a given point remains in the unstable vacuum is given by $P=e^{-I(T)}$, where~\cite{Guth:1979bh,Guth:1981uk}
\be
I(T) = \frac{4\pi}{3} \int_T^{T_c} \frac{{\rm d} T' \,\Gamma(T')}{T'^4 H(T')} \left( \int_{T}^{T'} \frac{ {\rm d} \tilde T}{H(\tilde T)} \right)^3 \,.
\ee
To find the percolation temperature, $T_*$, we solve for the condition $I(T_*) = 0.34$~\cite{Ellis:2018mja}, while ensuring that the volume of the false vacuum is decreasing:
\be
\frac{1}{V_{\rm false}}\frac{d V_{\rm false}}{dt}
=H(T)\left(3-T\frac{d I(T)}{d T}\right)<0 \, ,
\ee
so as to verify that percolation is possible despite the exponential expansion of the false vacuum.

The parameters relevant for the GW signal are then evaluated at $T=T_*$. These are the strength of the transition~\cite{Hindmarsh:2017gnf,Caprini:2019egz},
\be
\alpha \equiv \frac{1}{\rho_R} \left(\Delta V - \frac{T}{4} \Delta \frac{{\rm d}V}{{\rm d} T} \right)\bigg|_{T=T_*} \simeq \frac{\Delta V}{\rho_R(T_*)}\,,
\ee
and the characteristic length scale for the transition given by the average bubble radius~\cite{Turner:1992tz,Enqvist:1991xw},
\be
R_* =  \left[T_* \int_{T_*}^{T_c} \frac{{\rm d} T'}{T'^2} \frac{\Gamma(T')}{H(T')} e^{-I(T')} \right]^{-\frac13} \,,
\ee
which can be translated into the more commonly used time scale using $\beta=(8\pi)^\frac13/R_*$ for strong transitions.\footnote{In the case of a weak transition with the wall velocity dictated by strong interaction with the plasma, this expression needs a slight modification~\cite{Caprini:2019egz}.}

After the phase transition the scalar field oscillates around the true vacuum and decays heating the plasma. The decay of $\varphi$ is dominated by decays to two Higgs bosons, $\varphi \to hh$, and the decay width can be approximated as
\be \label{eq:Gammaphi}
\Gamma_\varphi \approx \frac{\lambda_p^2 v_\varphi^2}{32\pi m_\varphi} \approx 2.5\times 10^7 g_{B-L} \left(\frac{m_{Z'}}{\rm GeV}\right)^{-3} \,,
\ee
where we have used $m_\varphi \approx 0.4 g_{B-L} m_{Z'}$~\cite{Marzo:2018nov}. We discuss the possible matter-dominated period caused by slow decay of $\varphi$ and a suitably accurate prescription for the calculation of redshifting of the GW spectrum in this scenario in Section~\ref{sec:scaling}.

\begin{figure}
\centering 
\includegraphics[width=0.98\textwidth]{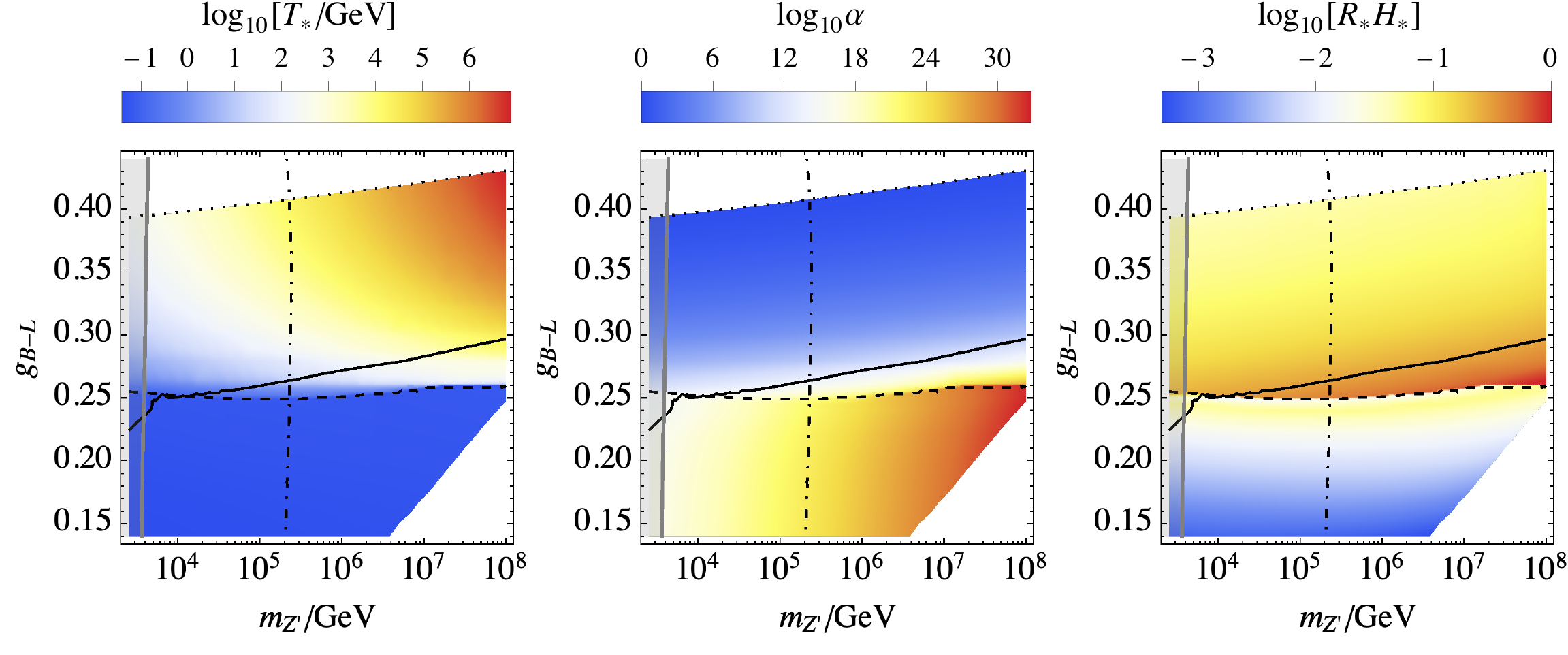}
\vspace{-3mm}
\caption{\it A selection of $(m_{Z'}, g_{B-L})$ planes, colour-coded according to the percolation temperature, $T_*$ (left), the strength of the transition, $\alpha$ (middle) and the average bubble radius at collisions, $R_*$ (right). Below the dotted curve $\alpha>1$, below the dashed curve $T_* < T_{\rm QCD}$, to the right of the dot-dashed curve $\Gamma_\phi < H_*$, and below the solid curve the bubbles collide before they reach terminal velocity (See Fig.~\ref{fig:gamma}). The region to the left of the gray line is excluded by collider searches.}
\label{fig:T}
\end{figure}

We display the phase transition parameters obtained in the classically conformal U$(1)_{B-L}$ model in Fig.~\ref{fig:T}, focusing on the region of parameter space where $\alpha>1$. We see that, in general, the percolation temperature $T_*$ {(left panel)} decreases and the strength of the transition $\alpha$ {(middle panel)} increases  for smaller values of the gauge coupling $g_{B-L}$. However, the trend in $\alpha$ changes when $T_*<T_{\rm QCD}$, at which point the transition is catalysed by quark condensation and the transition cannot become any stronger. For $T_*<T_{\rm QCD}$ the strength of the transition is determined by $m_{Z'}$. The average size of the bubbles at collision $R_*$ (right panel) is maximised compared to the horizon size when $g_{B-L} \simeq 0.25$, becoming slightly smaller for larger $g_{B-L}$ and significantly smaller for smaller $g_{B-L}$. We have cut off the lower right corners of the planes as there our assumption that $R_*$ is much bigger than the initial bubble radius does not hold (see Fig.~\ref{fig:gamma}).

The main modification due to the increase in $m_{Z'}$ is an increase in the vacuum energy difference, which we find to be related to the $Z'$ mass by
\be \label{eq:DeltaV}
\Delta V \approx 0.003 m_{Z'}^4 \,.
\ee
This causes a translation of the beginning of the vacuum-energy-dominated period, which happens at temperature  
\be
T_v = \frac{30\Delta V}{\pi^2 g_*(T_v)} \approx 0.1 m_{Z'} \,.
\ee
Using Eqs.~\eqref{eq:Gammaphi} and \eqref{eq:DeltaV}, we find the following approximate expression for $\alpha\gg 1$
\be
\label{GammaHRatio}
\frac{\Gamma_\varphi}{H_*} \approx 6 g_{B-L} \left(\frac{m_{Z'}}{2\times 10^5 {\rm GeV}}\right)^{-5} \,,
\ee
which shows that for high $Z'$ masses the $\varphi$ decay rate is smaller than the Hubble rate at the end of the transition. We indicate this by the dot-dashed curve in Fig.~\ref{fig:T} to the right of which $\Gamma_\varphi < H_*$.

\section{Bubble wall Lorentz factor and energy}
\label{sec:gamma}

\begin{figure}
\centering 
\includegraphics[width=0.98\textwidth]{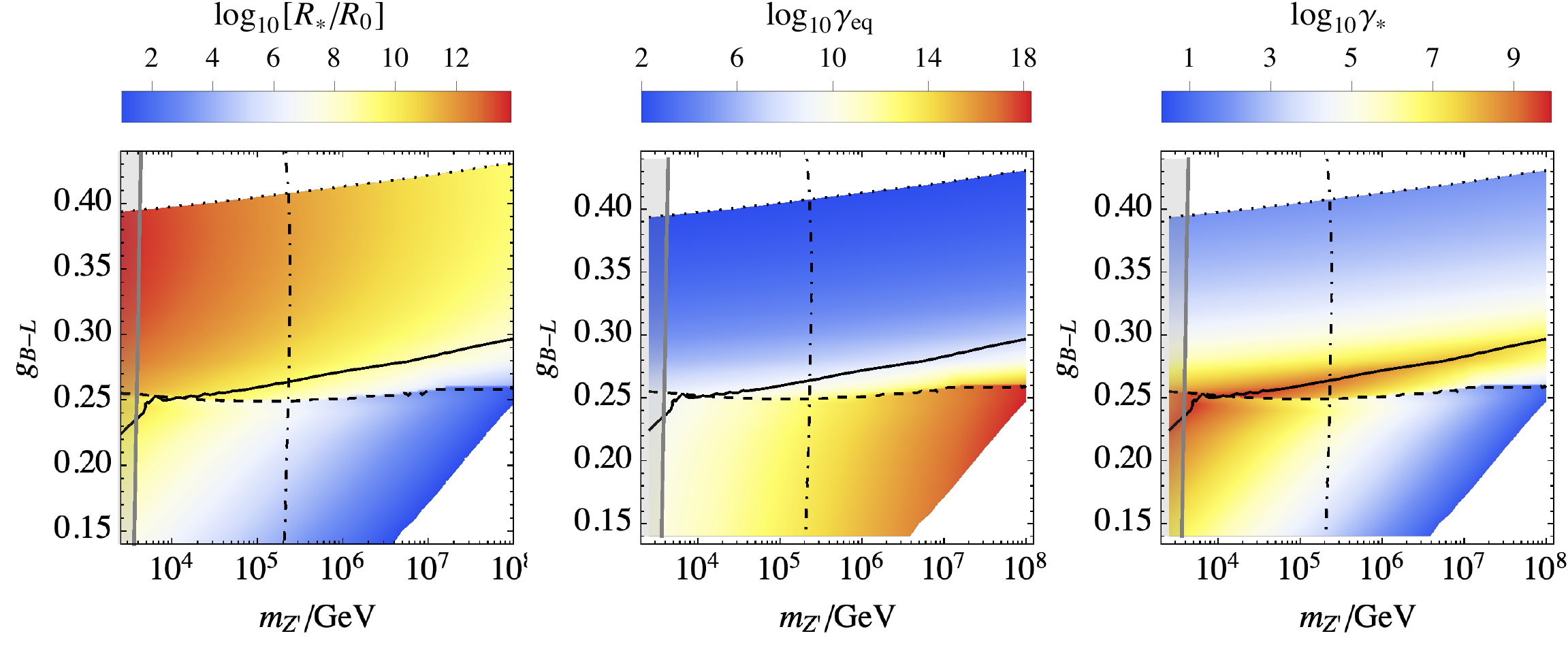}
\vspace{-3mm}
\caption{\it A selection of $(m_{Z'}, g_{B-L})$ planes, colour-coded according to the ratio of final and initial bubble radii, $R_*/R_0$ (left), the bubble wall Lorentz factor at which the equilibrium would be reached, $\gamma_{\rm eq}$ (middle) and the bubble wall Lorentz factor at the time of collisions, $\gamma_*$ (right). The black curves are the same as in Fig.~\ref{fig:T}.}
\label{fig:gamma}
\end{figure}

We next estimate the bubble wall Lorentz factor and whether the bubble wall reaches a terminal velocity before collisions. The pressure difference across the bubble wall is
\be
\Delta P = \Delta V - P_{1\to 1} - P_{1 \to N} \,,
\ee
where
\be
P_{1\to 1} \simeq 0.04 \Delta m^2 T_*^2\,, \qquad P_{1 \to N} \simeq 0.005 g^2 \gamma^2 T_*^4\,,
\ee
are the contributions from $1\to 1$ transitions~\cite{Bodeker:2009qy} and $1\to N$ splittings at the bubble wall~\cite{Hoeche:2020rsg} for $\gamma \gg 1$~\footnote{{The $1\to 2$ contribution discussed in~\cite{Bodeker:2017cim} is in principle included in the $1\to N$ contribution that is obtained by re-summing multiple scattering across the wall, as discussed in Ref.~\cite{Hoeche:2020rsg}}.}.
Here $\Delta m^2 = \sum_j c_j k_j \Delta m_j^2$ includes all particles whose mass changes in the transition, $g^2=\sum_j k_j g_j^2$ includes all gauge bosons to which the scalar field couples, and $\gamma$ denotes the Lorentz factor of the bubble wall. In equilibrium $\Delta P = 0$ and the corresponding bubble wall Lorentz factor is
\be
\gamma_{\rm eq} = \sqrt{\frac{\Delta V - P_{1\to 1}}{P_{1\to N}/\gamma^2}} \simeq \sqrt{\frac{\Delta V - 0.04\Delta m^2 T_*^2}{0.005g^2 T_*^4}}
\ee
in the classically conformal U$(1)_{B-L}$ model, where $\Delta m^2 \approx \Delta 3 m_{Z'}^2$ and $g^2 = 3 g_{B-L}^2$. We show $\gamma_{\rm eq}$ in the middle panel of Fig.~\ref{fig:gamma}. 

Neglecting the plasma effects, the bubble wall Lorentz factor increases as a function of the bubble radius as $\gamma \simeq 2R/(3R_0)$~\cite{Ellis:2019oqb}, where $R_0$ is the initial bubble radius. We calculate $R_0$ as described in Ref.~\cite{Ellis:2019oqb}. The left panel of Fig.~\ref{fig:gamma} shows how much the radius increases before the collisions. We have cut off the part of the parameter space where our calculation would give $R_* < 10 R_0$, as our calculation of $R_*$ is done assuming that $R_* \gg R_0$. At the time of collisions the bubble wall Lorentz factor is
\be \label{eq:gamma}
\gamma_* \simeq \min\left[\tilde\gamma_*, \gamma_{\rm eq}\right] \,, \quad {\rm where} \quad \tilde\gamma_* = \frac{2}{3} \frac{R_*}{R_0} \,.
\ee
As can be seen from the right panel of Fig.~\ref{fig:gamma}, the Lorentz factor can be very large, reaching $\gamma_* \simeq 10^9$ in the region where the collision time is similar to the time when the equilibrium is reached. Below the solid black line the bubbles collide before the terminal velocity is reached.

As discussed in Ref.~\cite{Ellis:2019oqb}, an important quantity for the GW calculation is the fraction of the total released energy that is in the scalar field gradients at the bubble wall. Here we improve the calculation of the bubble wall energy presented in Ref.~\cite{Ellis:2019oqb}. The bubble wall energy before collisions is
\be \label{eq:Ewall}
E_{\rm wall} = 4\pi R^2 \int_0^R \frac{\td R'}{3} \big[ \Delta V - P_{1\to 1} - P_{1\to N}(R') \big] \,,
\ee
where the integral gives the surface energy density of the bubble  and the prefactor is the area of the bubble surface. The second friction term arising from $1\to N$ splittings depends on the bubble radius through $P_{1\to N}\propto \gamma^2 = 4R^2/(9R_0^2)$. For $R>R(\gamma_{\rm eq})$ the integrand is zero and therefore the surface energy density remains constant after the terminal velocity is reached. By defining 
\be
\alpha_\infty = \frac{P_{1\to 1}}{\rho_R} \,,
\ee
we can write the ratio of the bubble wall energy and the total released energy, $E_V = 4\pi R^3 \Delta V/3$, which gives the so-called bubble collisions efficiency factor $\kappa_{\rm col}$, as
\be \label{eq:kappacol}
\kappa_{\rm col} = \frac{E_{\rm wall}}{E_V} = 
\begin{cases}
\bigg[1-\frac{1}{3}\left(\frac{\tilde\gamma_*}{\gamma_{\rm eq}}\right)^2\bigg] \bigg[1-\frac{\alpha_\infty}{\alpha}\bigg] \,, &\tilde\gamma_* < \gamma_{\rm eq} \,, \\
\frac{2}{3} \frac{\gamma_{\rm eq}}{\tilde\gamma_*} \bigg[1-\frac{\alpha_\infty}{\alpha}\bigg] \,, &\tilde\gamma_* > \gamma_{\rm eq} \,,
\end{cases}
\ee
where $\tilde\gamma_*$, defined in Eq.~\eqref{eq:gamma}, is the Lorentz factor that the wall would reach if the friction terms were neglected. Compared to Ref.~\cite{Ellis:2019oqb}, where the last term in the integrand~\eqref{eq:Ewall} was not accounted for, the main difference is the first factor in the case $\gamma_* < \gamma_{\rm eq}$, the absence of which caused a slight overestimation of the bubble wall energy in Ref.~\cite{Ellis:2019oqb}.

\section{Gravitational wave signal}
\label{sec:gws}

Next we discuss different GW sources in the phase transition. We will start with the GW signal produced by colliding bubbles. While the spectrum from this source can in principle be calculated in a 3D lattice simulation~\cite{Child:2012qg,Cutting:2018tjt,Cutting:2020nla}, practical realisations are made difficult by the hierarchy between the size of the growing bubble and its very thin wall in strong transitions. As a result it is practically impossible to simulate the strong transitions that would be necessary to produce this signal in realistic scenarios. Because of this, we use results from a simplified approach based on treating walls as infinitely thin shells~\cite{Kosowsky:1992vn}. Initially these walls were assumed simply to disappear upon collision as the energy dissipates~\cite{Huber:2008hg,Weir:2016tov,Konstandin:2017sat}. However, it was recently realised that a more realistic slower dissipation of the source will modify the signal~\cite{Jinno:2017fby,Konstandin:2017sat}. This leads us to the most recent result~\cite{Lewicki:2020jiv}, which uses the thin-shell approach to compute the GW spectrum while basing the dissipation of the sources on lattice simulations of two-bubble collisions. Following Ref.~\cite{Lewicki:2020jiv} we calculate the abundance of GWs sourced by the scalar field gradients as
\be
\Omega_{{\rm col},*} = 2.30\times 10^{-3}\left(R_* H_*\right)^{2}\left(\frac{\kappa_{\rm col} \alpha}{1+\alpha}\right)^{\!2} \!\left[1+\left(\frac{f}{f_{d}}\right)^{-1.61}\right]\!\left(\frac{f}{f_{{\rm col}}}\right)^{\!2.54}\!\left[1+1.13\left(\frac{f}{f_{{\rm col}}}\right)^{\!2.08}\right]^{-2.30}\!,
\ee
where the peak frequency is proportional to the inverse of the mean bubble separation,
\be
f_{{\rm col}} = 0.28 R_*^{-1} \,,
\ee
the frequency $f_{d}$ where the low-frequency slope changes is given by
\be
f_{d} = 0.044 R_*^{-1} \,,
\ee
and the efficiency factor $\kappa_{\rm col}$ is given in Eq.~\eqref{eq:kappacol}.

We turn next to GW sources related to motions in the primordial plasma, which are typically separated into two distinct periods. First there is the sound wave period observed in numerical lattice simulations involving the relativistic fluid and field bubbles~\cite{Hindmarsh:2013xza,Hindmarsh:2015qta,Hindmarsh:2017gnf}. This behaviour lasts until~\cite{Hindmarsh:2017gnf}
\be
\tau_{\rm sw} = \frac{R_*}{U_f} \,,
\ee
when the fluid motion becomes turbulent and the sound wave period ends~\footnote{This behaviour was recently also discussed in an expanding background~\cite{Guo:2020grp}. However, the modification is negligible for a sound wave period lasting much less than a Hubble time, which is what we find in the model studied here when the sound wave source is relevant. }. The root-mean- square fluid velocity $U_f$ can be approximated as~\cite{Hindmarsh:2015qta}
\be
U_f \simeq \sqrt{\frac34 \frac{\alpha_{\rm eff}}{1+\alpha_{\rm eff}} \kappa_{\rm sw}}\,,
\ee
where $\kappa_{\rm sw}$ is the efficiency coefficient for GW production from sound waves:
\be
\kappa_{\rm sw} =\frac{\alpha_{\rm eff}}{\alpha} \frac{\alpha_{\rm eff}}{0.73+0.083\sqrt{\alpha_{\rm eff}}+\alpha_{\rm eff}} \quad, \quad {\rm with} \,\,\,\,\, \alpha_{\rm eff} = \alpha(1-\kappa_{\rm col}) \,.
\ee
GWs are subsequently produced by the ensuing turbulence~\cite{Caprini:2009yp}. Following~\cite{Ellis:2019oqb}, we assume that at $\tau_{\rm sw}$ all the energy of the fluid motion is simply converted into turbulence, which implies a significant enhancement from this source compared to earlier estimates~\cite{Caprini:2015zlo}, as the sound wave period generically lasts much less than a Hubble time~\cite{Ellis:2018mja}. The final abundances of the plasma-related sources can be expressed as~\cite{Caprini:2015zlo,Ellis:2019oqb}
\bea
&\Omega_{{\rm sw},*} = 0.384 \, (\tau_{\rm sw}H_*) \left(R_* H_*\right) \left(\frac{\kappa_{\rm sw} \,\alpha }{1+\alpha }\right)^2 \left(\frac{f}{f_{\rm sw}}\right)^3 \left[1+\frac{3}{4} \left(\frac{f}{f_{\rm sw}}\right)^2\right]^{-\frac72} \,,\\
&\Omega_{{\rm turb},*} = 6.85 \,\left(R_* H_*\right) \left(1-\tau_{\rm sw}H_*\right) \left(\frac{\kappa_{\rm sw} \,\alpha }{1+\alpha }\right)^{3/2} \frac{\left(\frac{f}{f_{\rm turb}}\right)^3 \left[1+\left(\frac{f}{f_{\rm turb}}\right)\right]^{-\frac{11}{3}}}{1+8\pi f/H_*}\,,
\label{plasmasources}
\eea
with the peak frequencies given by
\bea
& f_{\rm sw} \,\simeq\, 3.4 R_*^{-1} \,,\\
& f_{\rm turb} \simeq 5.1 R_*^{-1} \,.
\eea
In order to compute the final spectra as observed today we also need the appropriate redshift factors, which we discuss at some length in the next Section.

We note that estimates of the plasma-related sources continuously undergo improvements, for instance through analytical description of plasma sound waves~\cite{Hindmarsh:2016lnk,Hindmarsh:2019phv}. In particular, the modelling of turbulence is far from settled, although much work has been invested in recent years~\cite{Caprini:2006jb,Kahniashvili:2009mf,Kahniashvili:2009qi,Kahniashvili:2012uj,Kahniashvili:2012vt,Niksa:2018ofa,Pol:2019yex,Ellis:2020uid}. Pending further developments in modelling these sources, we use the expressions in~\eqref{plasmasources} above as representative of the current state of the art most widely used in the community. 

\begin{figure}
\centering 
\includegraphics[width=0.98\textwidth]{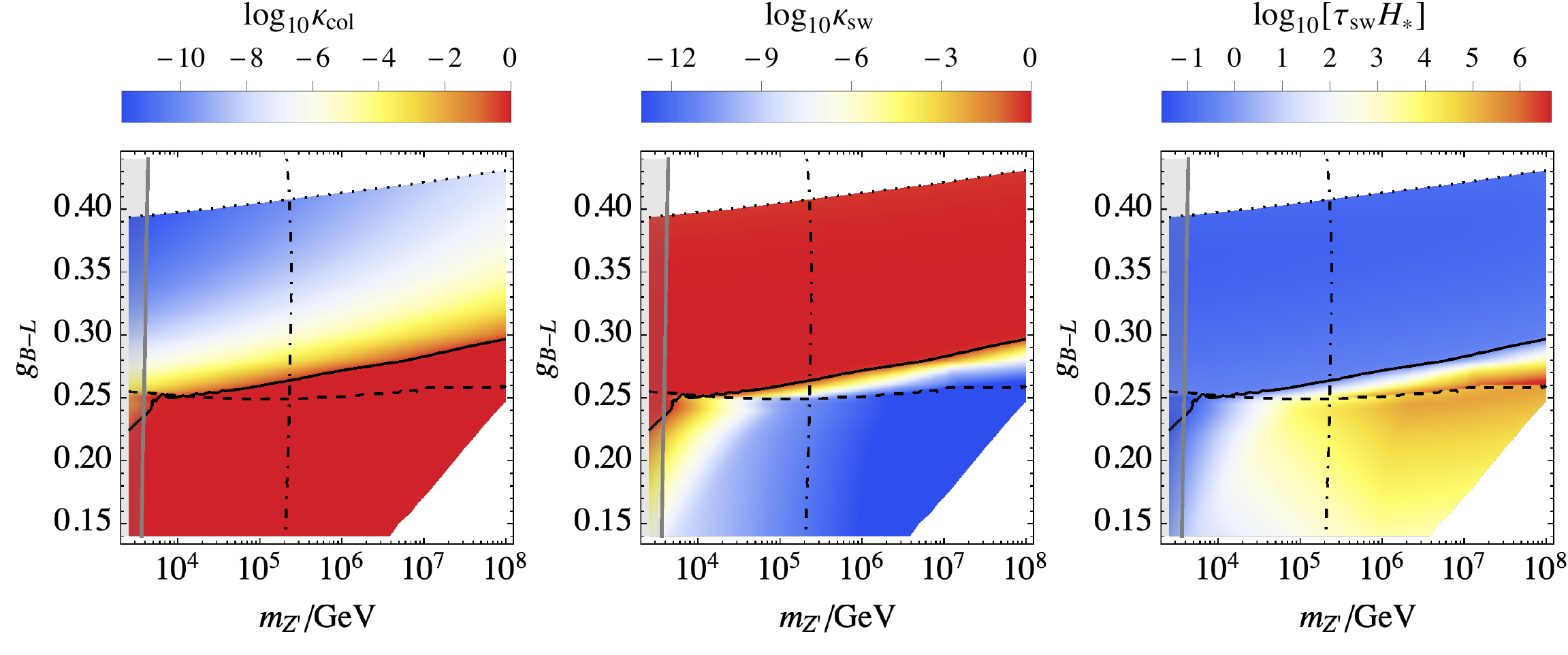}
\vspace{-3mm}
\caption{\it A selection of $(m_{Z'}, g_{B-L})$ planes, colour-coded according to the bubble collisions efficiency factor $\kappa_{\rm col}$, the sound wave efficiency factor $\kappa_{\rm sw}$ and the duration of the sound wave period $\tau_{\rm sw}$. The black curves are the same as in Fig.~\ref{fig:T}.}
\label{fig:kappa}
\end{figure}

We show in Fig.~\ref{fig:kappa} the efficiency factors $\kappa_{\rm col}$ {(left panel)} and $\kappa_{\rm sw}$ {(middle panel)}, and the duration of the sound wave period $\tau_{\rm sw}$ {(right panel)} in the classically conformal U$(1)_{B-L}$ model. We see that $\kappa_{\rm col} \sim 1$ below the solid black line where the bubble walls collide before reaching a terminal velocity. This indicates that the GW signal from the phase transition is dominantly sourced by the scalar field gradients. Above that line $\kappa_{\rm col}$ quickly drops and $\kappa_{\rm sw}\sim 1$. In this region the plasma-related sources dominate the GW production. As the duration of the sound wave period in this case is $\tau_{\rm sw} \sim 0.1 H_*^{-1}$, sound waves and turbulence in the plasma give comparable contributions to the GW signal.

\begin{figure}
\centering 
\includegraphics[width=10cm]{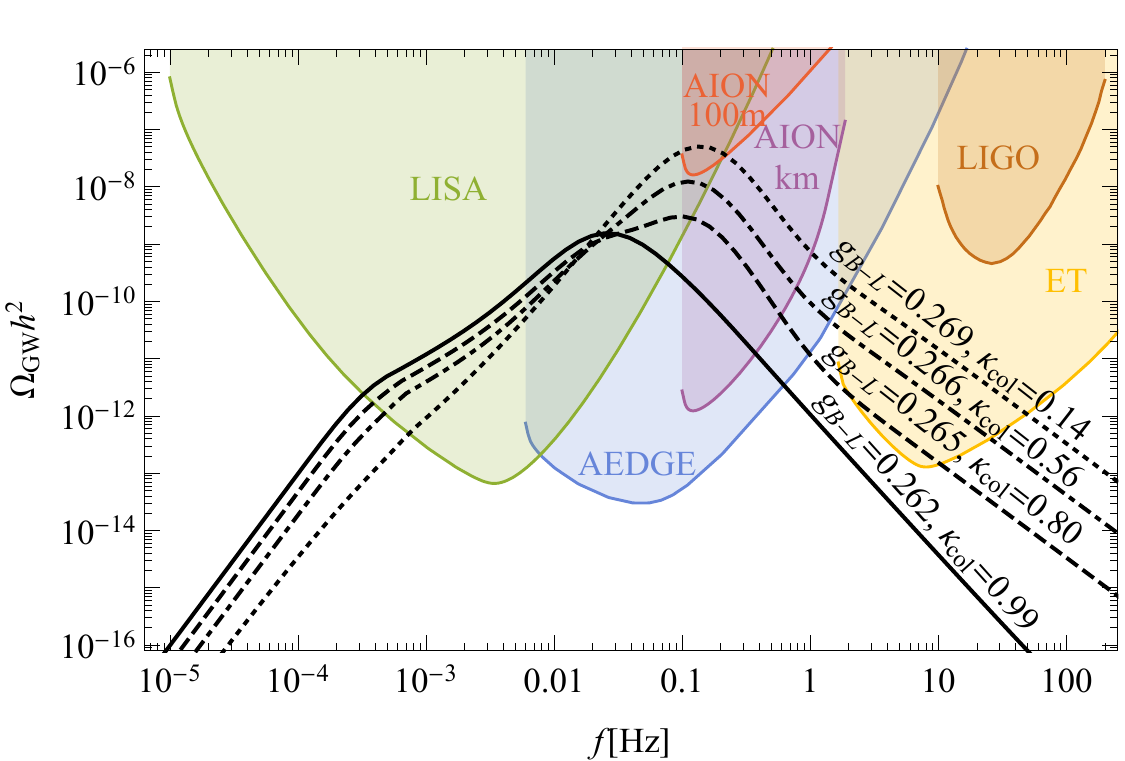}
\caption{\it Gravitational wave spectra generated in the U(1)$_{B-L}$ model for $m_{Z'} = 3.2\times 10^5\,$GeV and the range of indicated values of $g_{B-L}$. This range is chosen to exhibit the transition between the bubble collision signals and plasma signals as the gauge coupling increases.}
\label{fig:GWmod2}
\end{figure}

We illustrate the rapid transition from GW signals sourced by plasma motion to scalar field gradients in Fig.~\ref{fig:GWmod2}. To this end we set the U(1)$_{B - L}$ boson mass to $m_{Z'} = 3.2\times 10^5\,$GeV, and show results for several values of the gauge coupling in the range $0.262\leq g_{B-L} \leq 0.269$. This narrow range is enough to switch from signals dominated by plasma sources to a signal produced predominantly by bubble collisions. We also show the sensitivity of LIGO~\cite{TheLIGOScientific:2014jea,Thrane:2013oya,TheLIGOScientific:2016wyq}, as well as the projections for the future laser interferometers LISA~\cite{Bartolo:2016ami,Caprini:2019pxz} and Einstein Telescope~(ET)~\cite{Punturo:2010zz,Hild:2010id} and the terrestrial atom interferometer experiment AION~\cite{Badurina:2019hst}, its twin project MAGIS~\cite{Graham:2016plp,Graham:2017pmn}, and their proposed satellite version AEDGE~\cite{Bertoldi:2019tck}.

\section{Impact of matter domination on the GW signal}
\label{sec:scaling}

We turn now to a discussion of the redshift of the GW signal. We pay special attention to the case in which the very strong transition with $\alpha\gg 1$ is accompanied by very small field decay rate. This causes the field to oscillate around its minimum after the transition, before the Hubble rate drops below the field decay rate to finally allow reheating. If plasma effects can be neglected, these oscillations will dominate the energy density at the end of the transition. If instead the bubble walls reach a terminal velocity before collisions, the energy density after the transition is dominated by the plasma and scales as radiation. However, also in this case the field oscillations can become the dominant energy density component prior to their decay, causing a matter-dominated period.

Accounting for a fraction $1-\kappa_{\rm col}$ of the total released vacuum energy to go into the plasma during the expansion of the bubbles, the energy densities of the scalar field $\varphi$ and radiation at the time of the transition, $t=t_*$ are
\be \label{eq:Hfullinitial}
\rho_{\varphi*}=3 M_p^2 H_*^2 \frac{\kappa_{\rm col}\alpha}{1+\alpha} \,, \quad
\rho_{R*}=3 M_p^2 H_*^2 \frac{(1-\kappa_{\rm col})\,\alpha}{1+\alpha} \,. 
\ee
The subsequent decay of the scalar field oscillations into radiation is governed by the $\varphi$ decay rate, $\Gamma_\varphi$, and the evolution of the energy densities is governed by the equations
\be \label{eq:Hfull}
\dot{\rho}_R + 4H \rho_R = \Gamma_\varphi  \rho_\varphi \,, \quad
\dot{\rho}_\varphi + 3H \rho_\varphi = -\Gamma_\varphi  \rho_\varphi \,,\quad
H^2 = \frac{\rho_R+\rho_\varphi}{3 M_p^2}\, .
\ee
It is often convenient to rewrite these equations as functions of the scale factor $a$ rather than time, using $\td/\td t = aH \td/\td a$. We solve this set of equations numerically, finishing at the scale factor value $a=a_\times$ when the contribution of the field becomes negligible: $\rho_\varphi(a_{\times})/\rho_R(a_{\times})<10^{-3}\kappa_{\rm col}$. To stitch this result together with the subsequent standard radiation-dominated evolution, we simply convert the energy back to temperature via $\rho_R(a_{\times}) = \pi^2 g_*(T_{\times}) T_{\times}^4/30$,  assuming $g_*(T_\times)\approx 100$ for simplicity. 

\begin{figure}
\centering 
\includegraphics[width=10cm]{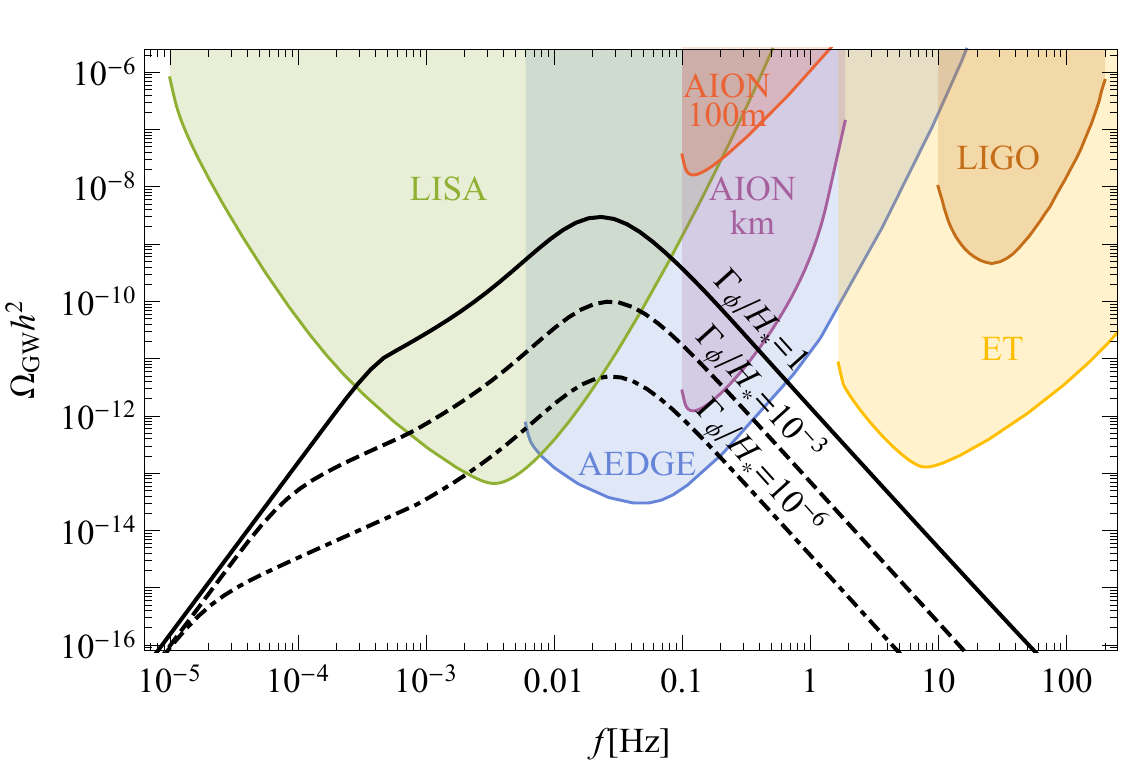}
\caption{\it GW spectra generated by bubble collisions for $g_{B-L} = 0.26$ and selected values of $m_{Z'}=3\times 10^{5}\,$GeV (solid) $m_{Z'}=8\times 10^{5}\,$GeV (dashed) $m_{Z'}=2\times 10^{6}\,$GeV (dot-dashed) in the U(1)$_{B-L}$ model. The corresponding values of $\Gamma_\varphi/H_*$ are also indicated above the signals. The period of matter-dominated expansion elongates the inclined `plateau' below the peak in the frequency spectrum, and also suppresses the peak as $m_{Z'}$ increases.}
\label{fig:GWmod1}
\end{figure}

Whether the field fluctuations decay quickly into radiation or not, the total energy during the transition is the same, and the initial abundance of GWs is not modified significantly~\cite{Allahverdi:2020bys}. However, after production the GWs redshift in the same way as radiation, and a period of matter domination modifies the GW abundance observed today. The GW spectrum redshifts as
\be \label{eq:PTOmegaredshift}
\Omega_{{\rm GW},0}(f) = 
\begin{cases}
\left(\frac{a_*}{a_0}\right)^4  \left(\frac{H_*}{H_0}\right)^2 \Omega_{{\rm GW},*}\!\left(\frac{a_0}{a_*} f\right) &{\rm for} \quad f>f_* \,, \\
\left(\frac{a_f}{a_0}\right)^4  \left(\frac{H_f}{H_0}\right)^2 \Omega_{{\rm GW},*}\!\left(\frac{a_0}{a_*} f_*\right) \left(\frac{f}{f_*}\right)^3 &{\rm for} \quad f<f_* \,,
\end{cases}
\ee
where $a_f$ corresponds to the scale factor when the scale $2\pi f$ re-enters the horizon, $2\pi f = a_f H_f$, and $2\pi f_* = a_* H_*$ corresponds to the horizon-size wavelength at the time of the phase transition. We can write the redshift factor for the total abundance as
\be 
\left(\frac{a}{a_0}\right)^4  \left(\frac{H}{H_0}\right)^2 = 1.67\times 10^{-5} h^{-2} \left(\frac{100}{g_{\rm eff}(T_{\times})}\right)^\frac13 \left(\frac{a}{a_{\times}}\right)^4  \left(\frac{H}{H_{\times}}\right)^2 \,,
\ee
and for the frequency as
\be
\frac{a_0}{a_*} = 2.70\times 10^{15} \left(\frac{T_\times}{100\,{\rm GeV}}\right) \left(\frac{g_{\rm eff}(T_{\times})}{100}\right)^\frac13 \frac{a_\times}{a_*} \,.
\ee
In addition to redshifting  the peak of the spectrum, another important effect is that a matter-dominated period can produce features in the spectrum, as superhorizon modes re-entering the horizon at different times get different redshift factors~\cite{Barenboim:2016mjm}. Beyond the horizon size at the time of production, the spectrum from a short-lasting source, such as a phase transition, is $\propto f^3$ in standard radiation-dominated expansion~\cite{Caprini:2009fx,Cai:2019cdl}, but the behaviour changes to $\propto f^1$ for scales re-entering the horizon during matter domination~\cite{Domenech:2020kqm}.
Therefore, during a matter-dominated period we observe a tilted `plateau' with slope $f^1$ developing below the frequency corresponding to the horizon size. The length of this plateau is dictated by the length of the matter-dominated period, which is given in our model by $\Gamma_\varphi/H_*$. We illustrate this effect in Fig.~\ref{fig:GWmod1} by showing the GW spectra for a few chosen points in the parameter space with different values of $\Gamma_\varphi/H_*$. As we see, for smaller decay rates the overall abundance of the spectrum is diminished across much of the spectrum by the modified redshifting, whereas the flatter `plateau' at small frequencies becomes longer. In this example, the `plateau' may be measurable by LISA, whereas the peak might be detectable by AEDGE as well as LISA.

This effect can be also described with a simple analytical approximation in the case that at $a=a_*$ the scalar field oscillations dominate the energy density, $\rho_{\varphi*} \gg \rho_{R*}$. We start by approximating to the solution of the system~\eqref{eq:Hfull} to describe the brief period of matter domination as~\cite{Barenboim:2016mjm}
\be \label{eq:Happrox}
H(a)=H_* \left(\frac{a_*}{a_{\rm reh}}\right)^\frac32 \left(\frac{a_{\rm reh}}{a}\right)^2 \left[ 1+\left(\frac{a_{\rm reh}}{a}\right)^2 \right]^{-\frac14}\, .
\ee
For this approximation to work properly it is crucial to find the value of the scale factor $a_{\rm reh}$ at which radiation domination resumes. Assuming that the expansion is dominated by field oscillations from $H_*$ we get  
\be
t - t_* = \int_{a_*}^a\frac{\td a}{a H} = \frac23\frac{1}{H_*} \left[\left(\frac{a}{a_*}\right)^\frac32 - 1 \right],
\ee
which, combined with the lifetime of the oscillations given by $\tau_\varphi=1/\Gamma_\varphi$, and assuming that $\Gamma_\varphi \ll H_*$, gives
\be \label{eq:areh_approx}
a_{\rm reh} = a_* \left( \frac32 \frac{H_*}{\Gamma_\varphi  } \right)^\frac23 \, .
\ee
In order to implement modifications to the GW signal we find the frequency corresponding to the horizon size at reheating, $f_{H {\rm reh}} = a_{\rm reh} H(a_{\rm reh})/2\pi$, and the frequency corresponding to the horizon size at transition, $f_*= a_{*} H(a_{*})/2\pi$. We can then use Eq.~\eqref{eq:Happrox} to solve for the evolution of the scale factor
\be
\frac{a}{a_{\rm reh}}=\sqrt{\frac{\chi(f)}{2}-\frac12}\, , \quad
\chi(f) \equiv \sqrt{8\left(\frac{f_{H {\rm reh}}}{f}\right)^4+1} \,.
\ee
The final step is to use this while taking care of the terms in Eq.~\eqref{eq:PTOmegaredshift} corresponding to redshifting up to $a_{\rm reh}$, obtaining
\be
\left(\frac{a}{a_{\rm reh}}\right)^4  \left(\frac{H}{H_{\rm reh}}\right)^2 \approx 
\begin{cases}
1 &{\rm for} \quad f<f_{H {\rm reh}} \,, \\
\sqrt{2\frac{\chi(f)-1}{\chi(f)+1}} & {\rm for} \quad f_{H {\rm reh}}<f<f_* \,, \\
\sqrt{2\frac{\chi(f_*)-1}{\chi(f_*)+1}} & {\rm for} \quad  f>f_* \,,
\end{cases}
\ee
where the frequency dependence and various cases take into account the redshifting of modes only after they re-enter the horizon. 

We have checked this approximation against our numerical results, finding that it is quite accurate provided that $\Gamma_{\varphi}/H_* \ll 1$ and $\kappa_{\rm col}\approx 1$, in which case it can easily be used to reproduce the $f^1$ plateau developing in the GW spectrum at low frequencies. However, we still use the full numerical result of Eq.~\eqref{eq:Hfull} in the results we show in the next Section, partly due to one additional caveat, namely that, even if $\kappa_{\rm col}$ in Eq.~\eqref{eq:Hfullinitial} is small and significant reheating occurs at the transition, the small amount of energy stored in field oscillations can still dominate the expansion for some time after the transition if the dominant radiation redshifts away before the field decay overtakes the expansion rate. In that case the additional plateau loses significance, as it is far detached from the peak, though the additional redshift can still make the spectra more difficult to observe.~\footnote{We could with effort find an analytical approximation for the Hubble rate also in that case, but the solution for the GW spectrum would be much more complicated, and would involve some numerics.}

\section{Prospects for observing the GW signals}
\label{sec:results}

\begin{figure}
\centering 
\includegraphics[width=7.2cm]{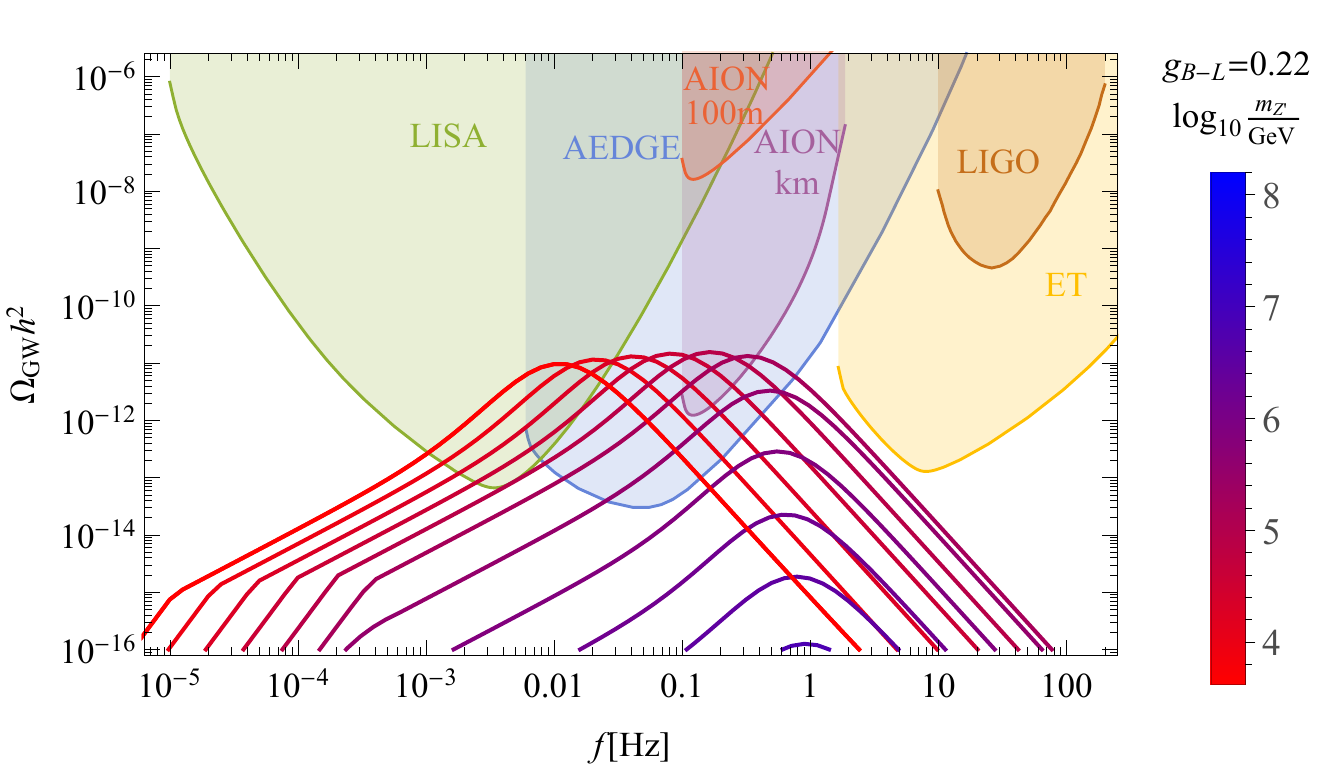}
\includegraphics[width=7.2cm]{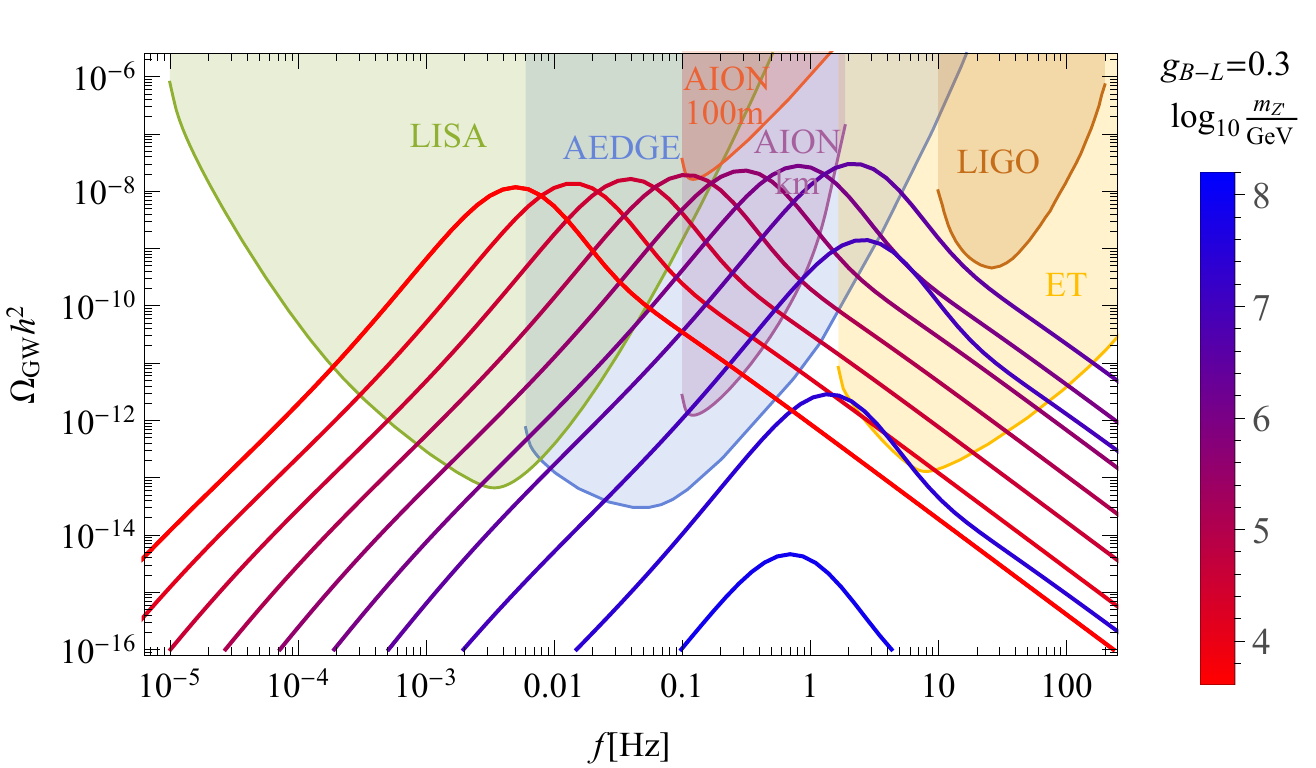}
\caption{\it GW spectra for selected values of $g_{B-L} = 0.22$ (left panel) and $0.3$ (right panel), and of $m_{Z'}$ (colour-coded) in the U(1)$_{B-L}$ model, calculated for the values of $\Gamma_\varphi/H_*$ given by Eq.~(\ref{GammaHRatio}). The period of matter-dominated expansion causes the tilted `plateau' below the peak in the frequency spectrum, and also the suppression of the peak for larger $m_{Z'}$.}
\label{fig:GWmodlines}
\end{figure}

We display in Fig.~\ref{fig:GWmodlines} the spectra of GWs for $g_{B-L} = 0.22$ (left panel) and $g_{B-L} = 0.30$ (right panel) and selected values of $m_{Z'}$ indicated by colour coding. Plasma sources of GWs such as sound waves and turbulence dominate the signal for $g_{B-L} = 0.3$, whereas for $g_{B-L} = 0.22$ the spectra are predominantly produced through bubble collisions. In both panels, the leftmost spectrum is that for the lightest $Z'$ boson consistent with LHC data, with a mass, $m_{Z'}\simeq 4\,$TeV. The damping of the signal visible at high masses is due to the modified redshifting of the signal caused by very slow decay of the field, which leads to a period of matter domination. This effect is more severe if initially most of the energy is stored in the scalar field profiles, as in the left panel, and we see that the signals are diminished before they reach ET frequencies. The same effect is visible in the right panel, although the matter-domination period there is postponed due to significant reheating during the transition and, as a result, higher peak frequencies are populated. The visible effect of this matter-domination period is the lengthening of the $\propto f^1$ `plateau' forming at frequencies below the peak in the lower spectra in the left panel. The same feature appears also in the right panel. However, because the matter-domination period begins significantly after the transition for this larger value of $g_{B-L}$, the feature appears much further from the peak and is not apparent in the figure.  
We see that when $g_{B-L} = 0.22$ in this model the LHC lower limit on $m_{Z'}$ reduces the chances for detection of a GW signal by LISA~\cite{Bartolo:2016ami,Caprini:2019pxz}, and that the suppression by matter domination of the peak for large $m_{Z'}$ appears to preclude measurement of this GW signal by ET~\cite{Punturo:2010zz,Hild:2010id}. On the other hand, AEDGE~\cite{Bertoldi:2019tck} appears to offer good prospects for GW detection for $m_{Z'} \lesssim 10^6\,$GeV and AION 1km~\cite{Badurina:2019hst} also has some prospects for $m_{Z'} \sim 10^6\,$TeV. On the other hand, when $g_{B-L} = 0.3$ each of LISA, AEDGE and ET has a good prospect for GW detection over some range of $m_{Z'}$, as does AION 1km, and even AION 100m has some prospects for $m_{Z'} \sim 10^6\,$TeV, though detection by LIGO~\cite{TheLIGOScientific:2014jea,Thrane:2013oya,TheLIGOScientific:2016wyq} appears difficult.

\begin{figure}
\centering
\includegraphics[width=0.99\textwidth]{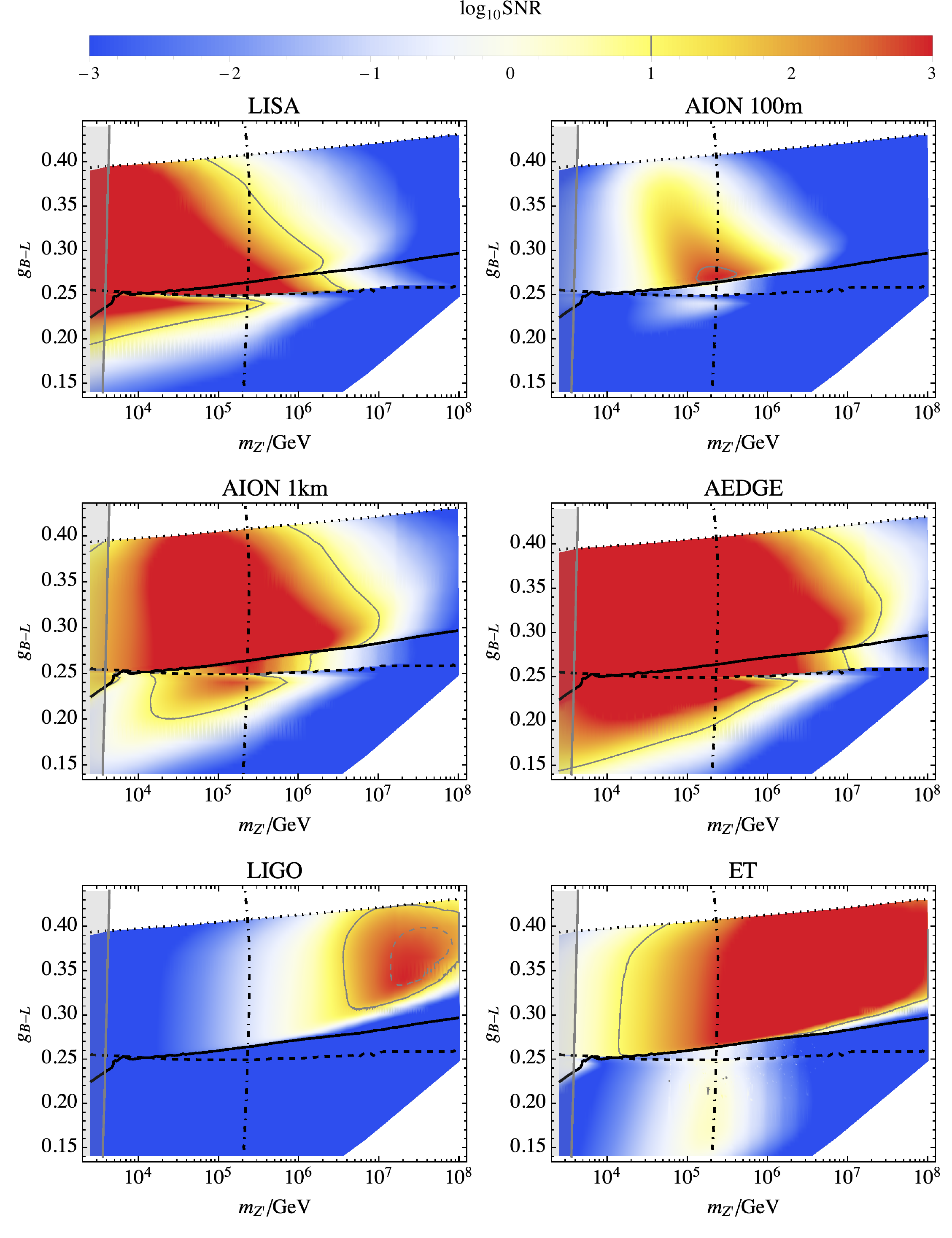}
\vspace{-3mm}
\caption{\it The expected signal-to-noise ratio (SNR) for observing the stochastic GW background from the U(1)$_{B-L}$ breaking phase transition for different future experiments: LISA (top left), AION 100m (top right), AION 1km (middle left), AEDGE (middle right), LIGO (bottom left),   and ET (bottom right). In the red regions ${\rm SNR} > 1000$ and in the blue regions ${\rm SNR} < 0.001$. The solid gray contours correspond to ${\rm SNR} = 10$, and the dashed gray contour in the bottom left panel indicates ${\rm SNR} = 1$ for the current LIGO sensitivity. Below the black solid curves the bubbles collide before they reach terminal velocity and the signal is dominated by the contribution from scalar field gradients, while above that line plasma-related sources dominate.}
\label{fig:SNR}
\end{figure}

Our next step is to probe systematically the parameter space for GW detection prospects in the various experiments. To this end, we make a standard signal-to-noise ratio (SNR) analysis. For each experiment we use its noise curve to calculate the SNR for the spectrum predicted at each point in the parameter space using
\be
{\rm SNR} \equiv \sqrt{\mathcal{T}\int {\rm d}f\, \left[\frac{\Omega_{\rm GW}(f)}{\Omega_{\rm noise}(f)}\right]^2} \,,
\ee
where we assume operation for a period $\mathcal{T}=4$~yr for each experiment. We display the results in Fig.~\ref{fig:SNR} for six experiments: LISA, AION 100m, AION 1km, AEDGE, LIGO and ET. {In the case of LIGO we show its design sensitivity. The current sensitivity does not reach ${\rm SNR}=10$, but the O2 sensitivity of LIGO reaches ${\rm SNR}>1$ within the region enclosed by the gray dashed contour shown in the upper right panel.} 

From Fig.~\ref{fig:SNR} we see that LISA {(top left panel)} has a good reach in the model parameter space, though the spectra in our model are typically pushed to frequencies above the peak sensitivity of LISA, due to collider constraints on the mass $m_{Z'}$ of the gauge boson. This means that the mid-frequency band is better suited to this model, and AEDGE and AION 1km {(middle panels)} can indeed probe a larger part of the allowed model parameter space. Below the solid black line the spectra are mostly produced by bubble collisions, whereas above it plasma-related sources dominate. We see that the effect of matter domination limiting the allowed frequencies from above is especially severe for bubble collision signals and, indeed, prevents ET {(bottom right panel)} from observing a bubble collision signal with significant SNR. This effect also diminishes the signal at high mass values, but other experiments, especially AEDGE, still have access to a significant part of the parameter space where this signal dominates. The situation is more standard for plasma-related sources where, as usual, larger mass scales are associated with higher frequencies, so LISA probes a lower mass range and ET a higher range, with AION and AEDGE covering the fullest range of frequencies.

\section{Conclusions}
\label{sec:concl}

We have revisited gravitational wave (GW) signals in the U(1)$_{B-L}$ extension of the Standard Model. We have probed the whole relevant parameter space, paying particular attention to regions with very strong transitions. We find that the GW signals are typically strong and should be detectable in upcoming experiments including LISA, AION/MAGIS, AEDGE, LIGO and ET. The frequency of the signal generically grows with the energy scale of the transition, which is in turn connected to the mass of the new U(1)$_{B-L}$ $Z'$ boson in the model. However, the optimal frequency range is limited from below by constraints on the mass of the the $Z'$ boson placed by LHC experiments, and
from above by the very slow decay of the U(1)$_{B-L}$-breaking field at high masses. This leads to a short matter-domination period that diminishes the signal.

We have refined the calculation of the energy budget for GW sources taking into account the scaling of the friction encountered by bubble walls as $\propto \gamma^2$. This means that the walls reach a terminal velocity faster than previously thought and, as a consequence, a smaller fraction of the total energy goes into accelerating the walls. Whilst this does modify the part of the parameter space predicting a strong bubble collision signal, we still find that a large part of the parameter space predicts a strong and potentially observable GW signal sourced predominantly by bubble collisions.  

We have scanned the parameter space of the model to identify where the GW signals produced would be within reach of upcoming experiments. We find that LISA as well as AION/MAGIS and AEDGE will be able to observe signals sourced mostly by bubble collision. Observation of such signals at higher frequencies is difficult as the large mass of the field prohibits its decay and triggers a matter-dominated epoch after the transition. This diminishes the signal, and we find that LIGO and ET will not be able to observe bubble collision spectra in our model with a large signal-to-noise ratio. For signals dominated by plasma sources this problem is mitigated, as significant reheating occurs already during the transition. However, even a small fraction of energy initially stored in field oscillations can eventually come to dominate due to its slower redshifting if the field decay is suppressed enough. Nevertheless, signals produced by plasma-related sources extend to much higher frequencies for appropriately large $Z'$ boson masses, and they would be observable by LIGO and ET as well as by LISA, AION/MAGIS and AEDGE.

\acknowledgments 
We are very grateful to Jos\'e~Miguel~No for useful discussions and carefully reading the manuscript. This work was supported by the UK STFC Grant ST/P000258/1. J.E. also acknowledges support from the Estonian Research Council grant MOBTT5, M.L. from the Polish National Science Center grant 2018/31/D/ST2/02048 and V.V. from the Estonian Research Council grant PRG803.

\bibliographystyle{JHEP}
\bibliography{PBH}

\providecommand{\href}[2]{#2}\begingroup\raggedright\begin{thebibliography}{100}

\bibitem{Kuzmin:1985mm}
V.~A. Kuzmin, V.~A. Rubakov and M.~E. Shaposhnikov, \emph{{On the Anomalous
  Electroweak Baryon Number Nonconservation in the Early Universe}},
  \href{https://doi.org/10.1016/0370-2693(85)91028-7}{\emph{Phys. Lett.}
  {\bfseries B155} (1985) 36}.

\bibitem{Cohen:1993nk}
A.~G. Cohen, D.~B. Kaplan and A.~E. Nelson, \emph{{Progress in electroweak
  baryogenesis}},
  \href{https://doi.org/10.1146/annurev.ns.43.120193.000331}{\emph{Ann. Rev.
  Nucl. Part. Sci.} {\bfseries 43} (1993) 27}
  [\href{https://arxiv.org/abs/hep-ph/9302210}{{\ttfamily hep-ph/9302210}}].

\bibitem{Riotto:1999yt}
A.~Riotto and M.~Trodden, \emph{{Recent progress in baryogenesis}},
  \href{https://doi.org/10.1146/annurev.nucl.49.1.35}{\emph{Ann. Rev. Nucl.
  Part. Sci.} {\bfseries 49} (1999) 35}
  [\href{https://arxiv.org/abs/hep-ph/9901362}{{\ttfamily hep-ph/9901362}}].

\bibitem{Morrissey:2012db}
D.~E. Morrissey and M.~J. Ramsey-Musolf, \emph{{Electroweak baryogenesis}},
  \href{https://doi.org/10.1088/1367-2630/14/12/125003}{\emph{New J. Phys.}
  {\bfseries 14} (2012) 125003}
  [\href{https://arxiv.org/abs/1206.2942}{{\ttfamily 1206.2942}}].

\bibitem{Abbott:2016blz}
{\scshape LIGO Scientific, Virgo} collaboration, \emph{{Observation of
  Gravitational Waves from a Binary Black Hole Merger}},
  \href{https://doi.org/10.1103/PhysRevLett.116.061102}{\emph{Phys. Rev. Lett.}
  {\bfseries 116} (2016) 061102}
  [\href{https://arxiv.org/abs/1602.03837}{{\ttfamily 1602.03837}}].

\bibitem{Witten:1984rs}
E.~Witten, \emph{{Cosmic Separation of Phases}},
  \href{https://doi.org/10.1103/PhysRevD.30.272}{\emph{Phys. Rev.} {\bfseries
  D30} (1984) 272}.

\bibitem{Kamionkowski:1993fg}
M.~Kamionkowski, A.~Kosowsky and M.~S. Turner, \emph{{Gravitational radiation
  from first order phase transitions}},
  \href{https://doi.org/10.1103/PhysRevD.49.2837}{\emph{Phys. Rev. D}
  {\bfseries 49} (1994) 2837}
  [\href{https://arxiv.org/abs/astro-ph/9310044}{{\ttfamily
  astro-ph/9310044}}].

\bibitem{Grojean:2006bp}
C.~Grojean and G.~Servant, \emph{{Gravitational Waves from Phase Transitions at
  the Electroweak Scale and Beyond}},
  \href{https://doi.org/10.1103/PhysRevD.75.043507}{\emph{Phys. Rev.}
  {\bfseries D75} (2007) 043507}
  [\href{https://arxiv.org/abs/hep-ph/0607107}{{\ttfamily hep-ph/0607107}}].

\bibitem{Espinosa:2008kw}
J.~Espinosa, T.~Konstandin, J.~No and M.~Quiros, \emph{{Some Cosmological
  Implications of Hidden Sectors}},
  \href{https://doi.org/10.1103/PhysRevD.78.123528}{\emph{Phys. Rev. D}
  {\bfseries 78} (2008) 123528}
  [\href{https://arxiv.org/abs/0809.3215}{{\ttfamily 0809.3215}}].

\bibitem{Dorsch:2014qpa}
G.~C. Dorsch, S.~J. Huber and J.~M. No, \emph{{Cosmological Signatures of a
  UV-Conformal Standard Model}},
  \href{https://doi.org/10.1103/PhysRevLett.113.121801}{\emph{Phys. Rev. Lett.}
  {\bfseries 113} (2014) 121801}
  [\href{https://arxiv.org/abs/1403.5583}{{\ttfamily 1403.5583}}].

\bibitem{Jaeckel:2016jlh}
J.~Jaeckel, V.~V. Khoze and M.~Spannowsky, \emph{{Hearing the signal of dark
  sectors with gravitational wave detectors}},
  \href{https://doi.org/10.1103/PhysRevD.94.103519}{\emph{Phys. Rev.}
  {\bfseries D94} (2016) 103519}
  [\href{https://arxiv.org/abs/1602.03901}{{\ttfamily 1602.03901}}].

\bibitem{Jinno:2016knw}
R.~Jinno and M.~Takimoto, \emph{{Probing a classically conformal B-L model with
  gravitational waves}},
  \href{https://doi.org/10.1103/PhysRevD.95.015020}{\emph{Phys. Rev. D}
  {\bfseries 95} (2017) 015020}
  [\href{https://arxiv.org/abs/1604.05035}{{\ttfamily 1604.05035}}].

\bibitem{Chala:2016ykx}
M.~Chala, G.~Nardini and I.~Sobolev, \emph{{Unified explanation for dark matter
  and electroweak baryogenesis with direct detection and gravitational wave
  signatures}}, \href{https://doi.org/10.1103/PhysRevD.94.055006}{\emph{Phys.
  Rev.} {\bfseries D94} (2016) 055006}
  [\href{https://arxiv.org/abs/1605.08663}{{\ttfamily 1605.08663}}].

\bibitem{Chala:2018opy}
M.~Chala, M.~Ramos and M.~Spannowsky, \emph{{Gravitational wave and collider
  probes of a triplet Higgs sector with a low cutoff}},
  \href{https://doi.org/10.1140/epjc/s10052-019-6655-1}{\emph{Eur. Phys. J.}
  {\bfseries C79} (2019) 156}
  [\href{https://arxiv.org/abs/1812.01901}{{\ttfamily 1812.01901}}].

\bibitem{Artymowski:2016tme}
M.~Artymowski, M.~Lewicki and J.~D. Wells, \emph{{Gravitational wave and
  collider implications of electroweak baryogenesis aided by non-standard
  cosmology}}, \href{https://doi.org/10.1007/JHEP03(2017)066}{\emph{JHEP}
  {\bfseries 03} (2017) 066}
  [\href{https://arxiv.org/abs/1609.07143}{{\ttfamily 1609.07143}}].

\bibitem{Hashino:2016xoj}
K.~Hashino, M.~Kakizaki, S.~Kanemura, P.~Ko and T.~Matsui, \emph{{Gravitational
  waves and Higgs boson couplings for exploring first order phase transition in
  the model with a singlet scalar field}},
  \href{https://doi.org/10.1016/j.physletb.2016.12.052}{\emph{Phys. Lett.}
  {\bfseries B766} (2017) 49}
  [\href{https://arxiv.org/abs/1609.00297}{{\ttfamily 1609.00297}}].

\bibitem{Vaskonen:2016yiu}
V.~Vaskonen, \emph{{Electroweak baryogenesis and gravitational waves from a
  real scalar singlet}},
  \href{https://doi.org/10.1103/PhysRevD.95.123515}{\emph{Phys. Rev. D}
  {\bfseries 95} (2017) 123515}
  [\href{https://arxiv.org/abs/1611.02073}{{\ttfamily 1611.02073}}].

\bibitem{Dorsch:2016nrg}
G.~Dorsch, S.~Huber, T.~Konstandin and J.~No, \emph{{A Second Higgs Doublet in
  the Early Universe: Baryogenesis and Gravitational Waves}},
  \href{https://doi.org/10.1088/1475-7516/2017/05/052}{\emph{JCAP} {\bfseries
  05} (2017) 052} [\href{https://arxiv.org/abs/1611.05874}{{\ttfamily
  1611.05874}}].

\bibitem{Beniwal:2017eik}
A.~Beniwal, M.~Lewicki, J.~D. Wells, M.~White and A.~G. Williams,
  \emph{{Gravitational wave, collider and dark matter signals from a scalar
  singlet electroweak baryogenesis}},
  \href{https://doi.org/10.1007/JHEP08(2017)108}{\emph{JHEP} {\bfseries 08}
  (2017) 108} [\href{https://arxiv.org/abs/1702.06124}{{\ttfamily
  1702.06124}}].

\bibitem{Baldes:2017rcu}
I.~Baldes, \emph{{Gravitational waves from the asymmetric-dark-matter
  generating phase transition}},
  \href{https://doi.org/10.1088/1475-7516/2017/05/028}{\emph{JCAP} {\bfseries
  05} (2017) 028} [\href{https://arxiv.org/abs/1702.02117}{{\ttfamily
  1702.02117}}].

\bibitem{Marzola:2017jzl}
L.~Marzola, A.~Racioppi and V.~Vaskonen, \emph{{Phase transition and
  gravitational wave phenomenology of scalar conformal extensions of the
  Standard Model}},
  \href{https://doi.org/10.1140/epjc/s10052-017-4996-1}{\emph{Eur. Phys. J.}
  {\bfseries C77} (2017) 484}
  [\href{https://arxiv.org/abs/1704.01034}{{\ttfamily 1704.01034}}].

\bibitem{Kang:2017mkl}
Z.~Kang, P.~Ko and T.~Matsui, \emph{{Strong first order EWPT \& strong
  gravitational waves in Z$_{3}$-symmetric singlet scalar extension}},
  \href{https://doi.org/10.1007/JHEP02(2018)115}{\emph{JHEP} {\bfseries 02}
  (2018) 115} [\href{https://arxiv.org/abs/1706.09721}{{\ttfamily
  1706.09721}}].

\bibitem{Iso:2017uuu}
S.~Iso, P.~D. Serpico and K.~Shimada, \emph{{QCD-Electroweak First-Order Phase
  Transition in a Supercooled Universe}},
  \href{https://doi.org/10.1103/PhysRevLett.119.141301}{\emph{Phys. Rev. Lett.}
  {\bfseries 119} (2017) 141301}
  [\href{https://arxiv.org/abs/1704.04955}{{\ttfamily 1704.04955}}].

\bibitem{Chala:2018ari}
M.~Chala, C.~Krause and G.~Nardini, \emph{{Signals of the electroweak phase
  transition at colliders and gravitational wave observatories}},
  \href{https://doi.org/10.1007/JHEP07(2018)062}{\emph{JHEP} {\bfseries 07}
  (2018) 062} [\href{https://arxiv.org/abs/1802.02168}{{\ttfamily
  1802.02168}}].

\bibitem{Bruggisser:2018mrt}
S.~Bruggisser, B.~Von~Harling, O.~Matsedonskyi and G.~Servant,
  \emph{{Electroweak Phase Transition and Baryogenesis in Composite Higgs
  Models}}, \href{https://doi.org/10.1007/JHEP12(2018)099}{\emph{JHEP}
  {\bfseries 12} (2018) 099}
  [\href{https://arxiv.org/abs/1804.07314}{{\ttfamily 1804.07314}}].

\bibitem{Megias:2018sxv}
E.~Megias, G.~Nardini and M.~Quiros, \emph{{Cosmological Phase Transitions in
  Warped Space: Gravitational Waves and Collider Signatures}},
  \href{https://doi.org/10.1007/JHEP09(2018)095}{\emph{JHEP} {\bfseries 09}
  (2018) 095} [\href{https://arxiv.org/abs/1806.04877}{{\ttfamily
  1806.04877}}].

\bibitem{Croon:2018erz}
D.~Croon, V.~Sanz and G.~White, \emph{{Model Discrimination in Gravitational
  Wave spectra from Dark Phase Transitions}},
  \href{https://doi.org/10.1007/JHEP08(2018)203}{\emph{JHEP} {\bfseries 08}
  (2018) 203} [\href{https://arxiv.org/abs/1806.02332}{{\ttfamily
  1806.02332}}].

\bibitem{Alves:2018jsw}
A.~Alves, T.~Ghosh, H.-K. Guo, K.~Sinha and D.~Vagie, \emph{{Collider and
  Gravitational Wave Complementarity in Exploring the Singlet Extension of the
  Standard Model}}, \href{https://doi.org/10.1007/JHEP04(2019)052}{\emph{JHEP}
  {\bfseries 04} (2019) 052}
  [\href{https://arxiv.org/abs/1812.09333}{{\ttfamily 1812.09333}}].

\bibitem{Baratella:2018pxi}
P.~Baratella, A.~Pomarol and F.~Rompineve, \emph{{The Supercooled Universe}},
  \href{https://doi.org/10.1007/JHEP03(2019)100}{\emph{JHEP} {\bfseries 03}
  (2019) 100} [\href{https://arxiv.org/abs/1812.06996}{{\ttfamily
  1812.06996}}].

\bibitem{Angelescu:2018dkk}
A.~Angelescu and P.~Huang, \emph{{Multistep Strongly First Order Phase
  Transitions from New Fermions at the TeV Scale}},
  \href{https://doi.org/10.1103/PhysRevD.99.055023}{\emph{Phys. Rev.}
  {\bfseries D99} (2019) 055023}
  [\href{https://arxiv.org/abs/1812.08293}{{\ttfamily 1812.08293}}].

\bibitem{Croon:2018kqn}
D.~Croon, T.~E. Gonzalo and G.~White, \emph{{Gravitational Waves from a
  Pati-Salam Phase Transition}},
  \href{https://doi.org/10.1007/JHEP02(2019)083}{\emph{JHEP} {\bfseries 02}
  (2019) 083} [\href{https://arxiv.org/abs/1812.02747}{{\ttfamily
  1812.02747}}].

\bibitem{Brdar:2018num}
V.~Brdar, A.~J. Helmboldt and J.~Kubo, \emph{{Gravitational Waves from
  First-Order Phase Transitions: LIGO as a Window to Unexplored Seesaw
  Scales}}, \href{https://doi.org/10.1088/1475-7516/2019/02/021}{\emph{JCAP}
  {\bfseries 1902} (2019) 021}
  [\href{https://arxiv.org/abs/1810.12306}{{\ttfamily 1810.12306}}].

\bibitem{Beniwal:2018hyi}
A.~Beniwal, M.~Lewicki, M.~White and A.~G. Williams, \emph{{Gravitational waves
  and electroweak baryogenesis in a global study of the extended scalar singlet
  model}}, \href{https://doi.org/10.1007/JHEP02(2019)183}{\emph{JHEP}
  {\bfseries 02} (2019) 183}
  [\href{https://arxiv.org/abs/1810.02380}{{\ttfamily 1810.02380}}].

\bibitem{Breitbach:2018ddu}
M.~Breitbach, J.~Kopp, E.~Madge, T.~Opferkuch and P.~Schwaller, \emph{{Dark,
  Cold, and Noisy: Constraining Secluded Hidden Sectors with Gravitational
  Waves}}, \href{https://doi.org/10.1088/1475-7516/2019/07/007}{\emph{JCAP}
  {\bfseries 07} (2019) 007}
  [\href{https://arxiv.org/abs/1811.11175}{{\ttfamily 1811.11175}}].

\bibitem{Marzo:2018nov}
C.~Marzo, L.~Marzola and V.~Vaskonen, \emph{{Phase transition and vacuum
  stability in the classically conformal B–L model}},
  \href{https://doi.org/10.1140/epjc/s10052-019-7076-x}{\emph{Eur. Phys. J.}
  {\bfseries C79} (2019) 601}
  [\href{https://arxiv.org/abs/1811.11169}{{\ttfamily 1811.11169}}].

\bibitem{Baldes:2018emh}
I.~Baldes and C.~Garcia-Cely, \emph{{Strong gravitational radiation from a
  simple dark matter model}},
  \href{https://doi.org/10.1007/JHEP05(2019)190}{\emph{JHEP} {\bfseries 05}
  (2019) 190} [\href{https://arxiv.org/abs/1809.01198}{{\ttfamily
  1809.01198}}].

\bibitem{Prokopec:2018tnq}
T.~Prokopec, J.~Rezacek and B.~Swiezewska, \emph{{Gravitational waves from
  conformal symmetry breaking}},
  \href{https://doi.org/10.1088/1475-7516/2019/02/009}{\emph{JCAP} {\bfseries
  1902} (2019) 009} [\href{https://arxiv.org/abs/1809.11129}{{\ttfamily
  1809.11129}}].

\bibitem{Fairbairn:2019xog}
M.~Fairbairn, E.~Hardy and A.~Wickens, \emph{{Hearing without seeing:
  gravitational waves from hot and cold hidden sectors}},
  \href{https://doi.org/10.1007/JHEP07(2019)044}{\emph{JHEP} {\bfseries 07}
  (2019) 044} [\href{https://arxiv.org/abs/1901.11038}{{\ttfamily
  1901.11038}}].

\bibitem{Helmboldt:2019pan}
A.~J. Helmboldt, J.~Kubo and S.~van~der Woude, \emph{{Observational prospects
  for gravitational waves from hidden or dark chiral phase transitions}},
  \href{https://doi.org/10.1103/PhysRevD.100.055025}{\emph{Phys. Rev.}
  {\bfseries D100} (2019) 055025}
  [\href{https://arxiv.org/abs/1904.07891}{{\ttfamily 1904.07891}}].

\bibitem{Dev:2019njv}
P.~S.~B. Dev, F.~Ferrer, Y.~Zhang and Y.~Zhang, \emph{{Gravitational Waves from
  First-Order Phase Transition in a Simple Axion-Like Particle Model}},
  \href{https://doi.org/10.1088/1475-7516/2019/11/006}{\emph{JCAP} {\bfseries
  1911} (2019) 006} [\href{https://arxiv.org/abs/1905.00891}{{\ttfamily
  1905.00891}}].

\bibitem{Jinno:2019jhi}
R.~Jinno, H.~Seong, M.~Takimoto and C.~M. Um, \emph{{Gravitational waves from
  first-order phase transitions: Ultra-supercooled transitions and the fate of
  relativistic shocks}},
  \href{https://doi.org/10.1088/1475-7516/2019/10/033}{\emph{JCAP} {\bfseries
  1910} (2019) 033} [\href{https://arxiv.org/abs/1905.00899}{{\ttfamily
  1905.00899}}].

\bibitem{Ellis:2019flb}
S.~A.~R. Ellis, S.~Ipek and G.~White, \emph{{Electroweak Baryogenesis from
  Temperature-Varying Couplings}},
  \href{https://doi.org/10.1007/JHEP08(2019)002}{\emph{JHEP} {\bfseries 08}
  (2019) 002} [\href{https://arxiv.org/abs/1905.11994}{{\ttfamily
  1905.11994}}].

\bibitem{Jinno:2019bxw}
R.~Jinno, T.~Konstandin and M.~Takimoto, \emph{{Relativistic bubble
  collisions—a closer look}},
  \href{https://doi.org/10.1088/1475-7516/2019/09/035}{\emph{JCAP} {\bfseries
  1909} (2019) 035} [\href{https://arxiv.org/abs/1906.02588}{{\ttfamily
  1906.02588}}].

\bibitem{Azatov:2019png}
A.~Azatov, D.~Barducci and F.~Sgarlata, \emph{{Gravitational traces of broken
  gauge symmetries}},
  \href{https://doi.org/10.1088/1475-7516/2020/07/027}{\emph{JCAP} {\bfseries
  07} (2020) 027} [\href{https://arxiv.org/abs/1910.01124}{{\ttfamily
  1910.01124}}].

\bibitem{Hindmarsh:2019phv}
M.~Hindmarsh and M.~Hijazi, \emph{{Gravitational waves from first order
  cosmological phase transitions in the Sound Shell Model}},
  \href{https://doi.org/10.1088/1475-7516/2019/12/062}{\emph{JCAP} {\bfseries
  1912} (2019) 062} [\href{https://arxiv.org/abs/1909.10040}{{\ttfamily
  1909.10040}}].

\bibitem{vonHarling:2019gme}
B.~Von~Harling, A.~Pomarol, O.~Pujolàs and F.~Rompineve, \emph{{Peccei-Quinn
  Phase Transition at LIGO}},
  \href{https://arxiv.org/abs/1912.07587}{{\ttfamily 1912.07587}}.

\bibitem{DelleRose:2019pgi}
L.~Delle~Rose, G.~Panico, M.~Redi and A.~Tesi, \emph{{Gravitational Waves from
  Supercool Axions}},
  \href{https://doi.org/10.1007/JHEP04(2020)025}{\emph{JHEP} {\bfseries 04}
  (2020) 025} [\href{https://arxiv.org/abs/1912.06139}{{\ttfamily
  1912.06139}}].

\bibitem{Vachaspati:1991nm}
T.~Vachaspati, \emph{{Magnetic fields from cosmological phase transitions}},
  \href{https://doi.org/10.1016/0370-2693(91)90051-Q}{\emph{Phys. Lett. B}
  {\bfseries 265} (1991) 258}.

\bibitem{Sigl:1996dm}
G.~Sigl, A.~V. Olinto and K.~Jedamzik, \emph{{Primordial magnetic fields from
  cosmological first order phase transitions}},
  \href{https://doi.org/10.1103/PhysRevD.55.4582}{\emph{Phys. Rev. D}
  {\bfseries 55} (1997) 4582}
  [\href{https://arxiv.org/abs/astro-ph/9610201}{{\ttfamily
  astro-ph/9610201}}].

\bibitem{Tevzadze:2012kk}
A.~G. Tevzadze, L.~Kisslinger, A.~Brandenburg and T.~Kahniashvili,
  \emph{{Magnetic Fields from QCD Phase Transitions}},
  \href{https://doi.org/10.1088/0004-637X/759/1/54}{\emph{Astrophys. J.}
  {\bfseries 759} (2012) 54} [\href{https://arxiv.org/abs/1207.0751}{{\ttfamily
  1207.0751}}].

\bibitem{Ellis:2019tjf}
J.~Ellis, M.~Fairbairn, M.~Lewicki, V.~Vaskonen and A.~Wickens,
  \emph{{Intergalactic Magnetic Fields from First-Order Phase Transitions}},
  \href{https://doi.org/10.1088/1475-7516/2019/09/019}{\emph{JCAP} {\bfseries
  1909} (2019) 019} [\href{https://arxiv.org/abs/1907.04315}{{\ttfamily
  1907.04315}}].

\bibitem{Moore:1995si}
G.~D. Moore and T.~Prokopec, \emph{{How fast can the wall move? A Study of the
  electroweak phase transition dynamics}},
  \href{https://doi.org/10.1103/PhysRevD.52.7182}{\emph{Phys. Rev.} {\bfseries
  D52} (1995) 7182} [\href{https://arxiv.org/abs/hep-ph/9506475}{{\ttfamily
  hep-ph/9506475}}].

\bibitem{Moore:1995ua}
G.~D. Moore and T.~Prokopec, \emph{{Bubble wall velocity in a first order
  electroweak phase transition}},
  \href{https://doi.org/10.1103/PhysRevLett.75.777}{\emph{Phys. Rev. Lett.}
  {\bfseries 75} (1995) 777}
  [\href{https://arxiv.org/abs/hep-ph/9503296}{{\ttfamily hep-ph/9503296}}].

\bibitem{Bodeker:2009qy}
D.~Bodeker and G.~D. Moore, \emph{{Can electroweak bubble walls run away?}},
  \href{https://doi.org/10.1088/1475-7516/2009/05/009}{\emph{JCAP} {\bfseries
  05} (2009) 009} [\href{https://arxiv.org/abs/0903.4099}{{\ttfamily
  0903.4099}}].

\bibitem{Kozaczuk:2015owa}
J.~Kozaczuk, \emph{{Bubble Expansion and the Viability of Singlet-Driven
  Electroweak Baryogenesis}},
  \href{https://doi.org/10.1007/JHEP10(2015)135}{\emph{JHEP} {\bfseries 10}
  (2015) 135} [\href{https://arxiv.org/abs/1506.04741}{{\ttfamily
  1506.04741}}].

\bibitem{Bodeker:2017cim}
D.~Bodeker and G.~D. Moore, \emph{{Electroweak Bubble Wall Speed Limit}},
  \href{https://doi.org/10.1088/1475-7516/2017/05/025}{\emph{JCAP} {\bfseries
  05} (2017) 025} [\href{https://arxiv.org/abs/1703.08215}{{\ttfamily
  1703.08215}}].

\bibitem{Dorsch:2018pat}
G.~C. Dorsch, S.~J. Huber and T.~Konstandin, \emph{{Bubble wall velocities in
  the Standard Model and beyond}},
  \href{https://doi.org/10.1088/1475-7516/2018/12/034}{\emph{JCAP} {\bfseries
  1812} (2018) 034} [\href{https://arxiv.org/abs/1809.04907}{{\ttfamily
  1809.04907}}].

\bibitem{Mancha:2020fzw}
M.~Barroso~Mancha, T.~Prokopec and B.~Swiezewska, \emph{{Field theoretic
  derivation of bubble wall force}},
  \href{https://arxiv.org/abs/2005.10875}{{\ttfamily 2005.10875}}.

\bibitem{Hoeche:2020rsg}
S.~Hoeche, J.~Kozaczuk, A.~J. Long, J.~Turner and Y.~Wang, \emph{{Towards an
  all-orders calculation of the electroweak bubble wall velocity}},
  \href{https://arxiv.org/abs/2007.10343}{{\ttfamily 2007.10343}}.

\bibitem{Ellis:2019oqb}
J.~Ellis, M.~Lewicki, J.~M. No and V.~Vaskonen, \emph{{Gravitational wave
  energy budget in strongly supercooled phase transitions}},
  \href{https://doi.org/10.1088/1475-7516/2019/06/024}{\emph{JCAP} {\bfseries
  1906} (2019) 024} [\href{https://arxiv.org/abs/1903.09642}{{\ttfamily
  1903.09642}}].

\bibitem{Allahverdi:2020bys}
R.~Allahverdi et~al., \emph{{The First Three Seconds: a Review of Possible
  Expansion Histories of the Early Universe}},
  \href{https://arxiv.org/abs/2006.16182}{{\ttfamily 2006.16182}}.

\bibitem{Barenboim:2016mjm}
G.~Barenboim and W.-I. Park, \emph{{Gravitational waves from first order phase
  transitions as a probe of an early matter domination era and its inverse
  problem}}, \href{https://doi.org/10.1016/j.physletb.2016.06.009}{\emph{Phys.
  Lett. B} {\bfseries 759} (2016) 430}
  [\href{https://arxiv.org/abs/1605.03781}{{\ttfamily 1605.03781}}].

\bibitem{TheLIGOScientific:2014jea}
{\scshape LIGO Scientific} collaboration, \emph{{Advanced LIGO}},
  \href{https://doi.org/10.1088/0264-9381/32/7/074001}{\emph{Class. Quant.
  Grav.} {\bfseries 32} (2015) 074001}
  [\href{https://arxiv.org/abs/1411.4547}{{\ttfamily 1411.4547}}].

\bibitem{Thrane:2013oya}
E.~Thrane and J.~D. Romano, \emph{{Sensitivity curves for searches for
  gravitational-wave backgrounds}},
  \href{https://doi.org/10.1103/PhysRevD.88.124032}{\emph{Phys. Rev.}
  {\bfseries D88} (2013) 124032}
  [\href{https://arxiv.org/abs/1310.5300}{{\ttfamily 1310.5300}}].

\bibitem{TheLIGOScientific:2016wyq}
{\scshape LIGO Scientific, Virgo} collaboration, \emph{{GW150914: Implications
  for the stochastic gravitational wave background from binary black holes}},
  \href{https://doi.org/10.1103/PhysRevLett.116.131102}{\emph{Phys. Rev. Lett.}
  {\bfseries 116} (2016) 131102}
  [\href{https://arxiv.org/abs/1602.03847}{{\ttfamily 1602.03847}}].

\bibitem{Bartolo:2016ami}
N.~Bartolo et~al., \emph{{Science with the space-based interferometer LISA. IV:
  Probing inflation with gravitational waves}},
  \href{https://doi.org/10.1088/1475-7516/2016/12/026}{\emph{JCAP} {\bfseries
  1612} (2016) 026} [\href{https://arxiv.org/abs/1610.06481}{{\ttfamily
  1610.06481}}].

\bibitem{Caprini:2019pxz}
C.~Caprini, D.~G. Figueroa, R.~Flauger, G.~Nardini, M.~Peloso, M.~Pieroni
  et~al., \emph{{Reconstructing the spectral shape of a stochastic
  gravitational wave background with LISA}},
  \href{https://doi.org/10.1088/1475-7516/2019/11/017}{\emph{JCAP} {\bfseries
  1911} (2019) 017} [\href{https://arxiv.org/abs/1906.09244}{{\ttfamily
  1906.09244}}].

\bibitem{Badurina:2019hst}
L.~Badurina et~al., \emph{{AION: An Atom Interferometer Observatory and
  Network}}, \href{https://doi.org/10.1088/1475-7516/2020/05/011}{\emph{JCAP}
  {\bfseries 05} (2020) 011}
  [\href{https://arxiv.org/abs/1911.11755}{{\ttfamily 1911.11755}}].

\bibitem{Graham:2016plp}
P.~W. Graham, J.~M. Hogan, M.~A. Kasevich and S.~Rajendran, \emph{{Resonant
  mode for gravitational wave detectors based on atom interferometry}},
  \href{https://doi.org/10.1103/PhysRevD.94.104022}{\emph{Phys. Rev.}
  {\bfseries D94} (2016) 104022}
  [\href{https://arxiv.org/abs/1606.01860}{{\ttfamily 1606.01860}}].

\bibitem{Graham:2017pmn}
{\scshape MAGIS} collaboration, \emph{{Mid-band gravitational wave detection
  with precision atomic sensors}},
  \href{https://arxiv.org/abs/1711.02225}{{\ttfamily 1711.02225}}.

\bibitem{Bertoldi:2019tck}
{\scshape AEDGE} collaboration, \emph{{AEDGE: Atomic Experiment for Dark Matter
  and Gravity Exploration in Space}},
  \href{https://doi.org/10.1140/epjqt/s40507-020-0080-0}{\emph{EPJ Quant.
  Technol.} {\bfseries 7} (2020) 6}
  [\href{https://arxiv.org/abs/1908.00802}{{\ttfamily 1908.00802}}].

\bibitem{Punturo:2010zz}
M.~Punturo et~al., \emph{{The Einstein Telescope: A third-generation
  gravitational wave observatory}},
  \href{https://doi.org/10.1088/0264-9381/27/19/194002}{\emph{Class. Quant.
  Grav.} {\bfseries 27} (2010) 194002}.

\bibitem{Hild:2010id}
S.~Hild et~al., \emph{{Sensitivity Studies for Third-Generation Gravitational
  Wave Observatories}},
  \href{https://doi.org/10.1088/0264-9381/28/9/094013}{\emph{Class. Quant.
  Grav.} {\bfseries 28} (2011) 094013}
  [\href{https://arxiv.org/abs/1012.0908}{{\ttfamily 1012.0908}}].

\bibitem{Iso:2009ss}
S.~Iso, N.~Okada and Y.~Orikasa, \emph{{Classically conformal B--L extended
  Standard Model}},
  \href{https://doi.org/10.1016/j.physletb.2009.04.046}{\emph{Phys. Lett. B}
  {\bfseries 676} (2009) 81} [\href{https://arxiv.org/abs/0902.4050}{{\ttfamily
  0902.4050}}].

\bibitem{Aaboud:2017buh}
{\scshape ATLAS} collaboration, \emph{{Search for new high-mass phenomena in
  the dilepton final state using 36 fb$^{−1}$ of proton-proton collision data
  at sqrt(s)=13 TeV with the ATLAS detector}},
  \href{https://doi.org/10.1007/JHEP10(2017)182}{\emph{JHEP} {\bfseries 10}
  (2017) 182} [\href{https://arxiv.org/abs/1707.02424}{{\ttfamily
  1707.02424}}].

\bibitem{Escudero:2018fwn}
M.~Escudero, S.~J. Witte and N.~Rius, \emph{{The dispirited case of gauged
  U(1)$_{B-L}$ dark matter}},
  \href{https://doi.org/10.1007/JHEP08(2018)190}{\emph{JHEP} {\bfseries 08}
  (2018) 190} [\href{https://arxiv.org/abs/1806.02823}{{\ttfamily
  1806.02823}}].

\bibitem{Wainwright:2011qy}
C.~Wainwright, S.~Profumo and M.~J. Ramsey-Musolf, \emph{{Gravity Waves from a
  Cosmological Phase Transition: Gauge Artifacts and Daisy Resummations}},
  \href{https://doi.org/10.1103/PhysRevD.84.023521}{\emph{Phys. Rev. D}
  {\bfseries 84} (2011) 023521}
  [\href{https://arxiv.org/abs/1104.5487}{{\ttfamily 1104.5487}}].

\bibitem{Chiang:2017zbz}
C.-W. Chiang and E.~Senaha, \emph{{On gauge dependence of gravitational waves
  from a first-order phase transition in classical scale-invariant $U(1)'$
  models}}, \href{https://doi.org/10.1016/j.physletb.2017.09.064}{\emph{Phys.
  Lett. B} {\bfseries 774} (2017) 489}
  [\href{https://arxiv.org/abs/1707.06765}{{\ttfamily 1707.06765}}].

\bibitem{Coriano:2015sea}
C.~Coriano, L.~Delle~Rose and C.~Marzo, \emph{{Constraints on abelian
  extensions of the Standard Model from two-loop vacuum stability and
  $U(1)_{B-L}$}}, \href{https://doi.org/10.1007/JHEP02(2016)135}{\emph{JHEP}
  {\bfseries 02} (2016) 135}
  [\href{https://arxiv.org/abs/1510.02379}{{\ttfamily 1510.02379}}].

\bibitem{Linde:1981zj}
A.~D. Linde, \emph{{Decay of the False Vacuum at Finite Temperature}},
  \href{https://doi.org/10.1016/0550-3213(83)90293-6,
  10.1016/0550-3213(83)90072-X}{\emph{Nucl. Phys.} {\bfseries B216} (1983)
  421}.

\bibitem{Ellis:2018mja}
J.~Ellis, M.~Lewicki and J.~M. No, \emph{{On the Maximal Strength of a
  First-Order Electroweak Phase Transition and its Gravitational Wave Signal}},
  \href{https://doi.org/10.1088/1475-7516/2019/04/003}{\emph{JCAP} {\bfseries
  04} (2019) 003} [\href{https://arxiv.org/abs/1809.08242}{{\ttfamily
  1809.08242}}].

\bibitem{Guth:1979bh}
A.~H. Guth and S.~H.~H. Tye, \emph{{Phase Transitions and Magnetic Monopole
  Production in the Very Early Universe}},
  \href{https://doi.org/10.1103/PhysRevLett.44.631,
  10.1103/PhysRevLett.44.963.2}{\emph{Phys. Rev. Lett.} {\bfseries 44} (1980)
  631}.

\bibitem{Guth:1981uk}
A.~H. Guth and E.~J. Weinberg, \emph{{Cosmological Consequences of a First
  Order Phase Transition in the SU(5) Grand Unified Model}},
  \href{https://doi.org/10.1103/PhysRevD.23.876}{\emph{Phys. Rev.} {\bfseries
  D23} (1981) 876}.

\bibitem{Hindmarsh:2017gnf}
M.~Hindmarsh, S.~J. Huber, K.~Rummukainen and D.~J. Weir, \emph{{Shape of the
  acoustic gravitational wave power spectrum from a first order phase
  transition}}, \href{https://doi.org/10.1103/PhysRevD.96.103520}{\emph{Phys.
  Rev.} {\bfseries D96} (2017) 103520}
  [\href{https://arxiv.org/abs/1704.05871}{{\ttfamily 1704.05871}}].

\bibitem{Caprini:2019egz}
C.~Caprini et~al., \emph{{Detecting gravitational waves from cosmological phase
  transitions with LISA: an update}},
  \href{https://doi.org/10.1088/1475-7516/2020/03/024}{\emph{JCAP} {\bfseries
  2003} (2020) 024} [\href{https://arxiv.org/abs/1910.13125}{{\ttfamily
  1910.13125}}].

\bibitem{Turner:1992tz}
M.~S. Turner, E.~J. Weinberg and L.~M. Widrow, \emph{{Bubble nucleation in
  first order inflation and other cosmological phase transitions}},
  \href{https://doi.org/10.1103/PhysRevD.46.2384}{\emph{Phys. Rev.} {\bfseries
  D46} (1992) 2384}.

\bibitem{Enqvist:1991xw}
K.~Enqvist, J.~Ignatius, K.~Kajantie and K.~Rummukainen, \emph{{Nucleation and
  bubble growth in a first order cosmological electroweak phase transition}},
  \href{https://doi.org/10.1103/PhysRevD.45.3415}{\emph{Phys. Rev.} {\bfseries
  D45} (1992) 3415}.

\bibitem{Child:2012qg}
H.~L. Child and J.~T. Giblin, \emph{{Gravitational Radiation from First-Order
  Phase Transitions}},
  \href{https://doi.org/10.1088/1475-7516/2012/10/001}{\emph{JCAP} {\bfseries
  10} (2012) 001} [\href{https://arxiv.org/abs/1207.6408}{{\ttfamily
  1207.6408}}].

\bibitem{Cutting:2018tjt}
D.~Cutting, M.~Hindmarsh and D.~J. Weir, \emph{{Gravitational waves from vacuum
  first-order phase transitions: from the envelope to the lattice}},
  \href{https://doi.org/10.1103/PhysRevD.97.123513}{\emph{Phys. Rev.}
  {\bfseries D97} (2018) 123513}
  [\href{https://arxiv.org/abs/1802.05712}{{\ttfamily 1802.05712}}].

\bibitem{Cutting:2020nla}
D.~Cutting, E.~G. Escartin, M.~Hindmarsh and D.~J. Weir, \emph{{Gravitational
  waves from vacuum first order phase transitions II: from thin to thick
  walls}},  \href{https://arxiv.org/abs/2005.13537}{{\ttfamily 2005.13537}}.

\bibitem{Kosowsky:1992vn}
A.~Kosowsky and M.~S. Turner, \emph{{Gravitational radiation from colliding
  vacuum bubbles: envelope approximation to many bubble collisions}},
  \href{https://doi.org/10.1103/PhysRevD.47.4372}{\emph{Phys. Rev.} {\bfseries
  D47} (1993) 4372} [\href{https://arxiv.org/abs/astro-ph/9211004}{{\ttfamily
  astro-ph/9211004}}].

\bibitem{Huber:2008hg}
S.~J. Huber and T.~Konstandin, \emph{{Gravitational Wave Production by
  Collisions: More Bubbles}},
  \href{https://doi.org/10.1088/1475-7516/2008/09/022}{\emph{JCAP} {\bfseries
  0809} (2008) 022} [\href{https://arxiv.org/abs/0806.1828}{{\ttfamily
  0806.1828}}].

\bibitem{Weir:2016tov}
D.~J. Weir, \emph{{Revisiting the envelope approximation: gravitational waves
  from bubble collisions}},
  \href{https://doi.org/10.1103/PhysRevD.93.124037}{\emph{Phys. Rev. D}
  {\bfseries 93} (2016) 124037}
  [\href{https://arxiv.org/abs/1604.08429}{{\ttfamily 1604.08429}}].

\bibitem{Konstandin:2017sat}
T.~Konstandin, \emph{{Gravitational radiation from a bulk flow model}},
  \href{https://doi.org/10.1088/1475-7516/2018/03/047}{\emph{JCAP} {\bfseries
  1803} (2018) 047} [\href{https://arxiv.org/abs/1712.06869}{{\ttfamily
  1712.06869}}].

\bibitem{Jinno:2017fby}
R.~Jinno and M.~Takimoto, \emph{{Gravitational waves from bubble dynamics:
  Beyond the Envelope}},
  \href{https://doi.org/10.1088/1475-7516/2019/01/060}{\emph{JCAP} {\bfseries
  01} (2019) 060} [\href{https://arxiv.org/abs/1707.03111}{{\ttfamily
  1707.03111}}].

\bibitem{Lewicki:2020jiv}
M.~Lewicki and V.~Vaskonen, \emph{{Gravitational wave spectra from strongly
  supercooled phase transitions}},
  \href{https://arxiv.org/abs/2007.04967}{{\ttfamily 2007.04967}}.

\bibitem{Hindmarsh:2013xza}
M.~Hindmarsh, S.~J. Huber, K.~Rummukainen and D.~J. Weir, \emph{{Gravitational
  waves from the sound of a first order phase transition}},
  \href{https://doi.org/10.1103/PhysRevLett.112.041301}{\emph{Phys. Rev. Lett.}
  {\bfseries 112} (2014) 041301}
  [\href{https://arxiv.org/abs/1304.2433}{{\ttfamily 1304.2433}}].

\bibitem{Hindmarsh:2015qta}
M.~Hindmarsh, S.~J. Huber, K.~Rummukainen and D.~J. Weir, \emph{{Numerical
  simulations of acoustically generated gravitational waves at a first order
  phase transition}},
  \href{https://doi.org/10.1103/PhysRevD.92.123009}{\emph{Phys. Rev.}
  {\bfseries D92} (2015) 123009}
  [\href{https://arxiv.org/abs/1504.03291}{{\ttfamily 1504.03291}}].

\bibitem{Guo:2020grp}
H.-K. Guo, K.~Sinha, D.~Vagie and G.~White, \emph{{Phase Transitions in an
  Expanding Universe: Stochastic Gravitational Waves in Standard and
  Non-Standard Histories}},  \href{https://arxiv.org/abs/2007.08537}{{\ttfamily
  2007.08537}}.

\bibitem{Caprini:2009yp}
C.~Caprini, R.~Durrer and G.~Servant, \emph{{The stochastic gravitational wave
  background from turbulence and magnetic fields generated by a first-order
  phase transition}},
  \href{https://doi.org/10.1088/1475-7516/2009/12/024}{\emph{JCAP} {\bfseries
  0912} (2009) 024} [\href{https://arxiv.org/abs/0909.0622}{{\ttfamily
  0909.0622}}].

\bibitem{Caprini:2015zlo}
C.~Caprini et~al., \emph{{Science with the space-based interferometer eLISA.
  II: Gravitational waves from cosmological phase transitions}},
  \href{https://doi.org/10.1088/1475-7516/2016/04/001}{\emph{JCAP} {\bfseries
  1604} (2016) 001} [\href{https://arxiv.org/abs/1512.06239}{{\ttfamily
  1512.06239}}].

\bibitem{Hindmarsh:2016lnk}
M.~Hindmarsh, \emph{{Sound shell model for acoustic gravitational wave
  production at a first-order phase transition in the early Universe}},
  \href{https://doi.org/10.1103/PhysRevLett.120.071301}{\emph{Phys. Rev. Lett.}
  {\bfseries 120} (2018) 071301}
  [\href{https://arxiv.org/abs/1608.04735}{{\ttfamily 1608.04735}}].

\bibitem{Caprini:2006jb}
C.~Caprini and R.~Durrer, \emph{{Gravitational waves from stochastic
  relativistic sources: Primordial turbulence and magnetic fields}},
  \href{https://doi.org/10.1103/PhysRevD.74.063521}{\emph{Phys. Rev.}
  {\bfseries D74} (2006) 063521}
  [\href{https://arxiv.org/abs/astro-ph/0603476}{{\ttfamily
  astro-ph/0603476}}].

\bibitem{Kahniashvili:2009mf}
T.~Kahniashvili, L.~Kisslinger and T.~Stevens, \emph{{Gravitational Radiation
  Generated by Magnetic Fields in Cosmological Phase Transitions}},
  \href{https://doi.org/10.1103/PhysRevD.81.023004}{\emph{Phys. Rev. D}
  {\bfseries 81} (2010) 023004}
  [\href{https://arxiv.org/abs/0905.0643}{{\ttfamily 0905.0643}}].

\bibitem{Kahniashvili:2009qi}
T.~Kahniashvili, A.~G. Tevzadze and B.~Ratra, \emph{{Phase Transition Generated
  Cosmological Magnetic Field at Large Scales}},
  \href{https://doi.org/10.1088/0004-637X/726/2/78}{\emph{Astrophys. J.}
  {\bfseries 726} (2011) 78} [\href{https://arxiv.org/abs/0907.0197}{{\ttfamily
  0907.0197}}].

\bibitem{Kahniashvili:2012uj}
T.~Kahniashvili, A.~G. Tevzadze, A.~Brandenburg and A.~Neronov,
  \emph{{Evolution of Primordial Magnetic Fields from Phase Transitions}},
  \href{https://doi.org/10.1103/PhysRevD.87.083007}{\emph{Phys. Rev. D}
  {\bfseries 87} (2013) 083007}
  [\href{https://arxiv.org/abs/1212.0596}{{\ttfamily 1212.0596}}].

\bibitem{Kahniashvili:2012vt}
T.~Kahniashvili, A.~Brandenburg, L.~Campanelli, B.~Ratra and A.~G. Tevzadze,
  \emph{{Evolution of inflation-generated magnetic field through phase
  transitions}}, \href{https://doi.org/10.1103/PhysRevD.86.103005}{\emph{Phys.
  Rev. D} {\bfseries 86} (2012) 103005}
  [\href{https://arxiv.org/abs/1206.2428}{{\ttfamily 1206.2428}}].

\bibitem{Niksa:2018ofa}
P.~Niksa, M.~Schlederer and G.~Sigl, \emph{{Gravitational Waves produced by
  Compressible MHD Turbulence from Cosmological Phase Transitions}},
  \href{https://doi.org/10.1088/1361-6382/aac89c}{\emph{Class. Quant. Grav.}
  {\bfseries 35} (2018) 144001}
  [\href{https://arxiv.org/abs/1803.02271}{{\ttfamily 1803.02271}}].

\bibitem{Pol:2019yex}
A.~Roper~Pol, S.~Mandal, A.~Brandenburg, T.~Kahniashvili and A.~Kosowsky,
  \emph{{Numerical Simulations of Gravitational Waves from Early-Universe
  Turbulence}},  \href{https://arxiv.org/abs/1903.08585}{{\ttfamily
  1903.08585}}.

\bibitem{Ellis:2020uid}
J.~Ellis, M.~Fairbairn, M.~Lewicki, V.~Vaskonen and A.~Wickens,
  \emph{{Detecting circular polarisation in the stochastic gravitational-wave
  background from a first-order cosmological phase transition}},
  \href{https://arxiv.org/abs/2005.05278}{{\ttfamily 2005.05278}}.

\bibitem{Caprini:2009fx}
C.~Caprini, R.~Durrer, T.~Konstandin and G.~Servant, \emph{{General Properties
  of the Gravitational Wave Spectrum from Phase Transitions}},
  \href{https://doi.org/10.1103/PhysRevD.79.083519}{\emph{Phys. Rev. D}
  {\bfseries 79} (2009) 083519}
  [\href{https://arxiv.org/abs/0901.1661}{{\ttfamily 0901.1661}}].

\bibitem{Cai:2019cdl}
R.-G. Cai, S.~Pi and M.~Sasaki, \emph{{Universal infrared scaling of
  gravitational wave background spectra}},
  \href{https://arxiv.org/abs/1909.13728}{{\ttfamily 1909.13728}}.

\bibitem{Domenech:2020kqm}
G.~Domènech, S.~Pi and M.~Sasaki, \emph{{Induced gravitational waves as a
  probe of thermal history of the universe}},
  \href{https://arxiv.org/abs/2005.12314}{{\ttfamily 2005.12314}}.

\end{thebibliography}\endgroup
\end{document}